\documentclass[draftclsnofoot,onecolumn]{IEEEtran}
\title{Joint Source-Channel Codes for MIMO Block Fading Channels}  % Declares the document's title.

\author{Deniz G\"{u}nd\"{u}z, Elza Erkip \thanks{The material in this paper was presented in
part at the 39th Asilomar Conference on Signals, Systems, and
Computers, Pacific Grove, CA, Nov. 2005, at the IEEE Information Theory Workshop,
Punta del Este, Uruguay, March 2006, and at the IEEE International Symposium on
Information Theory (ISIT), Seattle, WA, July 2006.}\thanks{This work is
partially supported by NSF grants No. 0430885 and No. 0635177.} \thanks{Deniz G\"{u}nd\"{u}z was with Department of Electrical and Computer Engineering, Polytechnic University. He is now with the Department of Electrical Engineering, Princeton University, Princeton, NJ, 08544, and also with the Department of Electrical Engineering, Stanford University, Stanford, CA, 94305. Elza Erkip is with the Department of Electrical and Computer Engineering, Polytechnic University, Brooklyn, NY, 11201. Email: dgunduz@princeton.edu, elza@poly.edu.}}
\date{}
\usepackage{amsfonts}
\usepackage{amssymb}
\usepackage{amsmath}
\usepackage{amscd}
\usepackage{graphicx}
\newcommand{\dst}{\displaystyle}

\newtheorem{thm}{Theorem}[section]
\newtheorem{cor}[thm]{Corollary}
\newtheorem{lem}[thm]{Lemma}
\newtheorem{defn}{Definition}[section]
\begin{document}
\fontsize{11}{13}\selectfont
% End of preamble and beginning of text.

\maketitle
\begin{abstract}

We consider transmission of a continuous amplitude source over
an $L$-block Rayleigh fading $M_t\times M_r$ MIMO channel when the
channel state information is only available at the receiver. Since the channel is not ergodic, Shannon's
source-channel separation theorem becomes obsolete and the optimal
performance requires a joint source -channel
approach. Our goal is to minimize the expected end-to-end distortion,
particularly in the high SNR regime. The
figure of merit is the distortion exponent, defined as the exponential decay
rate of the expected distortion with increasing SNR. We provide an
upper bound and lower bounds for the distortion exponent with
respect to the bandwidth ratio among the channel and source bandwidths.
For the lower bounds, we analyze three different strategies based on
layered source coding concatenated with progressive, superposition
or hybrid digital/analog transmission. In each case, by adjusting
the system parameters we optimize the distortion exponent as a
function of the bandwidth ratio. We prove that the distortion
exponent upper bound can be achieved when the channel has only one
degree of freedom, that is $L=1$, and $\min\{M_t,M_r\}=1$. When we
have more degrees of freedom, our achievable distortion exponents
meet the upper bound for only certain ranges of the bandwidth ratio. We
demonstrate that our results, which were derived for a complex
Gaussian source, can be extended to more general source
distributions as well.

\end{abstract}


\section{Introduction}\label{s:introduction}

\PARstart{R}{ecent} advances in mobile computing and hardware technology enable transmission of rich multimedia contents over wireless
networks. Examples include digital TV, voice and video
transmission over cellular and wireless LAN networks, and sensor
networks. With the high
demand for such services, it becomes crucial to identify the system limitations, define the
appropriate performance metrics, and to design wireless systems
that are capable of achieving the best performance by overcoming the
challenges posed by the system requirements and the wireless
environment. In general, multimedia wireless communication requires
transmitting analog sources over fading channels while satisfying the
end-to-end average distortion and delay requirements of the
application within the power limitations of the mobile terminal.

Multiple antennas at the transceivers have been proposed as a viable
tool that can remarkably improve the performance of multimedia
transmission over wireless channels. The additional degrees of freedom provided by multiple input multiple output (MIMO) system can be
utilized in the form of spatial multiplexing gain and/or spatial
diversity gain, that is, either to transmit more information or to
increase the reliability of the transmission. The tradeoff between
these two gains is explicitly characterized as the
diversity-multiplexing gain tradeoff (DMT) in \cite{Tse}. How to
translate this tradeoff into an improved overall system performance
depends on the application requirements and limitations.

In this paper, we consider transmission of a continuous amplitude
source over a MIMO block Rayleigh fading
channel. We are interested in minimizing the end-to-end average
distortion of the source. We assume that the instantaneous channel state information
is only available at the receiver (CSIR). We consider the case where $K$ source samples are to be transmitted
over $L$ fading blocks spanning $N$ channel uses. We
define the \textit{bandwidth ratio} of the system as
\begin{equation}\label{bexp}
\mathit{b}=\frac{N}{K} \mbox{ channel uses per source sample} ,
\end{equation}
and analyze the system performance with respect to $\textit{b}$. We
assume that $K$ is large enough to achieve the rate-distortion
performance of the underlying source, and $N$ is large enough to
design codes that can achieve all rates below the instantaneous
capacity of the block fading channel.

We are particularly interested in the high $SNR$ behavior of the
expected distortion (ED) which is characterized by the
\emph{distortion exponent} \cite{Emin}:
\begin{equation}\label{d:dist_exp}
\Delta=-\lim_{SNR\rightarrow\infty}\frac{\log ED}{\log SNR}.
\end{equation}

Shannon's fundamental source-channel separation theorem does not
apply to our scenario as the channel is no more ergodic. Thus, the
optimal strategy requires a joint source-channel coding approach.
The minimum expected end-to-end distortion depends on the source
characteristics, the distortion metric, the power
constraint of the transmitter, the joint compression, channel coding
and transmission techniques used.

Since we are interested in the average distortion of the system,
this requires a strategy that performs `well' over a range of
channel conditions. Our approach is to first compress the
source into multiple layers, where each layer successively refines
the previous layers, and then transmit these layers at varying
rates, hence providing unequal error protection so that the
reconstructed signal quality can be adjusted to the instantaneous
fading state without the availability of the channel state information at
the transmitter (CSIT). We consider transmitting the source layers either
progressively in time, \emph{layered source coding with progressive
transmission} (LS), or simultaneously by superposition,
\emph{broadcast strategy with layered source} (BS). We also discuss
a hybrid digital-analog extension of LS called
\emph{hybrid-LS }(HLS).

The characterization of distortion
exponent for fading channels has recently been investigated in
several papers. Distortion exponent is first defined in \cite{Emin}, and simple transmission schemes over two parallel fading channels are compared in terms of distortion exponent. Our prior work includes maximizing distortion exponent using layered source transmission for cooperative relay \cite{ISIT04, SPAWC, relayJ}, for  SISO
\cite{Asil}, for MIMO \cite{ITW}, and for parallel channels \cite{ISIT06}. Holliday and Goldsmith \cite{Holliday} analyze high SNR behavior of the expected distortion for single layer transmission over MIMO without explicitly giving the achieved distortion exponent. Hybrid digital-analog transmission, first proposed in
\cite{Phamdo} for the Gaussian broadcast channel, is considered in terms of distortion exponent for MIMO channel in \cite{Caire}.

Others have focused on minimizing the end-to-end source distortion
for general SNR values \cite{Honig, Sesia}. Recently, the LS and BS
strategies introduced here have been analyzed
for finite SNR and finite number of source layers in
\cite{Farzad1}-\cite{Tian}.

This paper derives explicit expressions for the achievable distortion exponent of LS,
HLS and BS strategies, and compares the achievable exponents with an
upper bound derived by assuming perfect channel state information at
the transmitter. Our results reveal the following:

\begin{itemize}
\item LS strategy, which can easily be implemented by concatenating
a layered source coder with a MIMO channel encoder that time-shares among different code rates,
improves the distortion exponent compared to the single-layer approach of
\cite{Holliday} even with limited number of source layers. However,
the distortion exponent of LS still falls short of the upper bound.

\item While the hybrid digital-analog scheme meets the distortion exponent
upper bound for small bandwidth ratios as shown in \cite{Caire}, the improvement of hybrid extension of LS (HLS) over pure progressive layered digital transmission (LS) becomes insignificant as the bandwidth ratio or the number of digital layers increases.

\item Transmitting layers simultaneously as suggested by BS provides
the optimal distortion exponent for all bandwidth ratios when the
system has one degree of freedom, i.e., for single block
MISO/SIMO systems, and for high bandwidth ratios for the general
MIMO system. Hence, for the mentioned cases the problem of characterizing the distortion
exponent is solved.

\item There is a close relationship between the DMT of the underlying MIMO channel and the achievable
distortion exponent of the proposed schemes. For LS and HLS, we are
able to give an explicit characterization of the achievable
distortion exponent once the DMT of the system is provided. For BS, we enforce successive decoding at the receiver and the achievable distortion exponent closely relates to the `\emph{successive decoding diversity-multiplexing tradeoff}' which will be rigorously defined in Section \ref{s:BS}.

\item The correspondence between source transmission to a single user with
unknown noise variance and multicasting to users with different noise levels
\cite{Phamdo} suggests that, our analysis would also apply to the
multicasting case where each receiver has the same number of antennas
and observes an independent block fading Rayleigh channel possibly with a different mean. Here the goal is to
minimize the expected distortion of each user. Alternatively, each
user may have a static channel, but the channel gains over the users may be randomly distributed
with independent Rayleigh distribution, where the objective is to minimize
the source distortion averaged over the users.

\item While minimizing the end-to-end distortion for finite SNR is still an open problem, in the high SNR regime we are able to provide a complete solution in certain scenarios. Using this high SNR analysis, it is also
possible to generalize the results to non-Gaussian source
distributions. Furthermore, LS and BS
strategies motivate source and channel coding strategies that are
shown to perform very well for finite SNRs as well \cite{Farzad1}-\cite{Tian}.

\end{itemize}

We use the following notation throughout the paper. $E[\cdot]$ is
the expectation, $f(x)\doteq g(x)$ is the exponential equality
defined as $\lim_{x\rightarrow\infty}\frac{\log f(x)}{\log g(x)}=1$, while $\dot{\geq}$ and $\dot{\leq}$ are defined similarly.
Vectors and matrices are denoted with bold characters, where matrices are in capital letters. $[\cdot]^T$
and $[\cdot]^\dag$ are the transpose and the conjugate transpose
operations, respectively. $\mathbf{tr}(\mathbf{A})$ is the the trace
of matrix $\mathbf{A}$. For two Hermitian matrices
$\mathbf{A}\succeq \mathbf{B}$ means that
$\mathbf{A}-\mathbf{B}$ is positive-semidefinite. $(x)^+$ is $x$ if
$x\geq 0$, and $0$ otherwise. We denote the set
$\{[x_1,\ldots,x_n]:x_i\in \mathbb{R}_+, \forall i\}$ by
$\mathbb{R}^{n+}$.

\section{System Model}\label{s:system}

We consider a discrete time continuous amplitude (analog) source
$\{s_k\}_{k=1}^\infty$, $s_k \in \mathbb{R}$ available at the
transmitter. For the analysis, we focus on a memoryless, i.i.d.,
complex Gaussian source with independent real and imaginary
components each with variance 1/2. We use the distortion-rate
function of the complex Gaussian source $D(R)=2^{-R}$ where $R$ is
the source coding rate in bits per source sample, and consider
compression strategies that meet the distortion-rate bound. Although
in Sections \ref{s:UB}-\ref{s:block_fading} we use properties of
this complex Gaussian source (such as its distortion-rate function
and successive refinability), in Section \ref{sectgen} we prove that
our results can be extended to any complex source with finite second
moment and finite differential entropy, with squared-error
distortion metric. As stated in Section \ref{s:introduction}, we
assume that $K$ source samples are transmitted in $N$ channel uses
which corresponds to a bandwidth ratio of $\textit{b}=N/K$. In all
our derivations we allow for an arbitrary bandwidth ratio
$\mathit{b}>0$.

We assume a MIMO block fading channel with $M_t$ transmit and
$M_r$ receive antennas. The channel model is
\begin{equation}\label{channel}
\mathbf{y}[i]=\sqrt{\frac{SNR}{M_t}}\mathbf{H}[i]\mathbf{x}[i]+\mathbf{z}[i],
\hspace{.3in} i=1,\ldots,N
\end{equation}
where $\sqrt{\frac{SNR}{M_t}}\mathbf{x}[i]$ is the transmitted signal at time $i$,
$\mathbf{Z}=[\mathbf{z}_1,\ldots,\mathbf{z}_N]\in\mathbb{C}^{M_r\times N}$
is the complex Gaussian noise with i.i.d entries
$\mathcal{CN}(0,1)$, and $\mathbf{H}[i]\in \mathbb{C}^{M_r\times
M_t}$ is the channel matrix at time $i$, which has i.i.d. entries
with $\mathcal{CN}(0,1)$. We have an $L$-block fading channel, that is, the channel observes $L$ different i.i.d.
fading realizations $\mathbf{H}_1,\ldots,\mathbf{H}_L$ each lasting
for $N/L$ channel uses. Thus we have
\begin{equation}
\mathbf{H}\left[k\frac{N}{L}+1\right]=\mathbf{H}\left[k\frac{N}{L}+2\right]=\ldots
=\mathbf{H}\left[(k+1)\frac{N}{L}\right]=\mathbf{H}_{k+1},
\end{equation}
for $k=0,\ldots,L-1$ assuming $N/L$ is integer. The realization of
the channel matrix $\mathbf{H}_i$ is assumed to be known by the
receiver and unknown by the transmitter, while the transmitter knows
the statistics of $\mathbf{H}_i$. The codeword, $\mathbf{X}=\bigg[
\mathbf{x}[1], \ldots, \mathbf{x}_L \bigg]\in\mathbb{C}^{M_t\times
N}$ is normalized so that it satisfies
$tr(E[\mathbf{X}^\dag\mathbf{X}])\leq M_tN$. We assume Gaussian codebooks which can
achieve the instantaneous capacity of the MIMO channel. We define
$M_*=\min\{M_t,M_r\}$ and $M^*=\max\{M_t,M_r\}$.

The source is transmitted through the channel using one of the joint
source-channel coding schemes discussed in this paper. In general,
the source encoder matches the $K$-length source vector
$\mathbf{s}^K=[s_1,\ldots, s_K]$ to the channel input $\mathbf{X}$.
The decoder maps the received signal
$\mathbf{Y}=\bigg[\mathbf{y}[1],\ldots,\mathbf{y}[N]\bigg]\in\mathbb{C}^{M_r\times
N}$ to an estimate $\hat{\mathbf{s}}\in\mathbb{C}^K$ of the source.
Average distortion $ED(SNR)$ is defined as the average mean squared
error between $\mathbf{s}$ and $\hat{\mathbf{s}}$ at average channel
signal-to-noise ratio $SNR$, where the expectation is taken with
respect to all source samples, channel realizations and the channel
noise. The exact expression of $ED(SNR)$ for the strategies
introduced will be provided in the respective sections.

As mentioned in Section \ref{s:introduction}, we are interested in
the high SNR behavior of the expected distortion. We optimize the
system performance to maximize the \textit{distortion exponent}
defined in Eqn. (\ref{d:dist_exp}). A distortion exponent of $\Delta$
means that the expected distortion decays as $SNR^{-\Delta}$ with
increasing SNR when $SNR$ is high.

In order to obtain the end-to-end distortion for our proposed
strategies, we will need to characterize the error rate of the MIMO
channel. Since we are interested in the high SNR regime, we use the
outage probability, which has the same exponential
behavior as the channel error probability \cite{Tse}. For a family
of codes with rate $R=r\log SNR$, $r$ is defined as the multiplexing
gain of the family, and
\begin{equation}
d(r)=- \lim_{SNR\rightarrow\infty}\frac{\log P_{out}(SNR)}{\log SNR}
\end{equation}
as the diversity advantage, where $P_{out}(SNR)$ is the outage probability of the code. The diversity gain $d^*(r)$ is defined
as the supremum of the diversity advantage over all possible code
families with multiplexing gain $r$. In \cite{Tse}, it is shown
that there is a fundamental tradeoff between multiplexing and
diversity gains, also known as the diversity-multiplexing gain tradeoff (DMT), and this tradeoff is explicitly characterized with
the following theorem.

\begin{thm}(Corollary 8, \cite{Tse})\label{divmimo}
For an $M_t\times M_r$ MIMO $L$-block fading channel, the optimal
tradeoff curve $d^*(r)$ is given by the piecewise-linear function
connecting the points $(k,d^*(k))$, $k=0,1,\ldots,M_*$, where
\begin{equation}
d^*(k)=L(M_t-k)(M_r-k).
\end{equation}
\end{thm}

\section{Distortion Exponent Upper Bound}\label{s:UB}

Before we study the performance of various source-channel coding
strategies, we calculate an upper bound for the distortion exponent of
the MIMO $L$-block fading channel, assuming that the transmitter has access to perfect channel
state information at the beginning of each block. Then
the source-channel separation theorem applies for each block and transmission at the highest rate is possible with zero outage probability.

\begin{thm}\label{t:upperbound}
For transmission of a memoryless i.i.d. complex Gaussian source over an $L$-block $M_t \times M_r$ MIMO
channel, the distortion exponent is upper bounded by
\begin{equation}
\Delta^{UB}=L\sum_{i=1}^{M_*}\min\left\{\frac{\textit{b}}{L},
2i-1+|M_t-M_r|\right\}.
\end{equation}\vspace{.01in}
\end{thm}
\begin{proof}
Proof of the theorem can be found in Appendix
\ref{APP:UB}.
\end{proof}

Note that, increasing the number of antennas at either the transmitter or the receiver by one does not provide
an increase in the distortion exponent upper bound for $\textit{b}<L(1+|M_t-M_r|)$,
since the performance in this region is bounded by the bandwidth
ratio. Adding one antenna to both sides increases the upper bound for all
bandwidth ratios, while the increase is more pronounced for higher
bandwidth ratios. The distortion exponent is
bounded by the highest diversity gain $LM_tM_r$.

In the case of $M\times 1$ MISO system, and alternatively $1\times
M$ SIMO system, the upper bound can be simplified to
\begin{eqnarray}
\Delta^{UB}_{MISO/SIMO} = \min\{\textit{b}, LM\}. \label{UB_MISO}
\end{eqnarray}

We next discuss how a simple transmission strategy consisting of single layer digital transmission
performs with respect to the upper bound. In single layer digital
transmission, the source is first compressed at a specific rate
$\textit{b}R$, the compressed bits are channel coded at rate $R$,
and then transmitted over the channel. This is the approach taken in
\cite{Emin} for two-parallel channels, in \cite{ISIT04} for cooperative relay channels, and in \cite{Holliday} for the
MIMO channel to transmit an analog source over a fading channel.
Even though compression and channel coding are done separately, the
rate is a common parameter that can be chosen to minimize the
end-to-end distortion. Note that the transmitter choses this
rate $R$ without any channel state information.

The expected
distortion of single layer transmission can be written as
\begin{equation}\label{e:motivation}
ED(R,SNR)=(1-P_{out}(R,SNR))D(\textit{b}R)+P_{out}(R,SNR),
\end{equation}
where $P_{out}(R,SNR)$ is the outage probability at rate $R$ for
given $SNR$, and $D(R)$ is the distortion-rate function of the
source. Here we assume that, in case of an outage, the decoder simply outputs the mean of the source leading to
the highest possible distortion of $1$ due to the unit variance
assumption. At fixed SNR, there is a tradeoff between reliable
transmission over the channel (through the outage probability), and
increased fidelity in source reconstruction (through the
distortion-rate function). This suggests that there is an optimal
transmission rate that achieves the optimal average distortion. For any
given SNR this optimal $R$ can be found using the exact expressions
for $P_{out}(R,SNR)$ and $D(R)$.

%---------------------------
\begin{figure}
\centering
\includegraphics[width=3in]{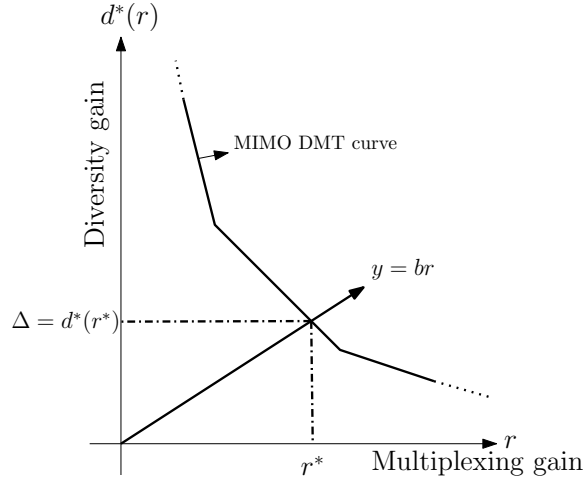}
\caption{A geometric interpretation illustrating the
optimal multiplexing gain for a single layer source-channel coding
system. The intersection point of the DMT curve
and the line \emph{y=br} gives the optimal multiplexing gain-
distortion exponent pair.} \label{single_layer}
\end{figure}
%---------------------------

In order to study the distortion exponent, we concentrate on the
high SNR approximation of Eqn. (\ref{e:motivation}). To achieve a
vanishing expected distortion in the high SNR regime we need to
increase $R$ with SNR. Scaling $R$ faster than $O(\log SNR)$ would
result in an outage probability of $1$, since the instantaneous
channel capacity of the MIMO system scales as $M_*\log SNR$. Thus we
assign $R=r\log SNR$, where $0\leq r\leq M_*$. Then the high SNR
approximation of Eqn. (\ref{e:motivation}) is
\begin{eqnarray}
ED &\doteq& D(\textit{b}R) + P_{out}(R), \nonumber \\
  &\doteq& SNR^{-\mathit{b}r} + SNR ^{-d^*(r)}.
\end{eqnarray}

Of the two terms, the one with the highest SNR exponent would be
dominant in the high SNR regime. Maximum distortion exponent is
achieved when both terms have the same SNR exponent. Then the
optimal multiplexing gain $r^*$ satisfies
\begin{eqnarray}\label{e:delta_sl}
\Delta \triangleq br^* = d^*(r^*),
\end{eqnarray}
where $\Delta$ is the corresponding distortion
exponent. Eqn. (\ref{e:delta_sl}) suggests an optimal operating
point on the DMT curve to maximize the distortion exponent of the single layer scheme.

Figure \ref{single_layer} shows a geometric illustration of the
optimal multiplexing gain and the corresponding distortion exponent.
A similar approach was taken in \cite{Holliday} for single layer transmission with the
restriction of integer multiplexing gains, and later extended to all
multiplexing gains in \cite{Caire}. However, as we argue next, even
when all multiplexing gains are considered, this
single layer approach is far from exploiting all the resources
provided by the system.

\begin{figure}
\centering
\includegraphics[width=3.5in]{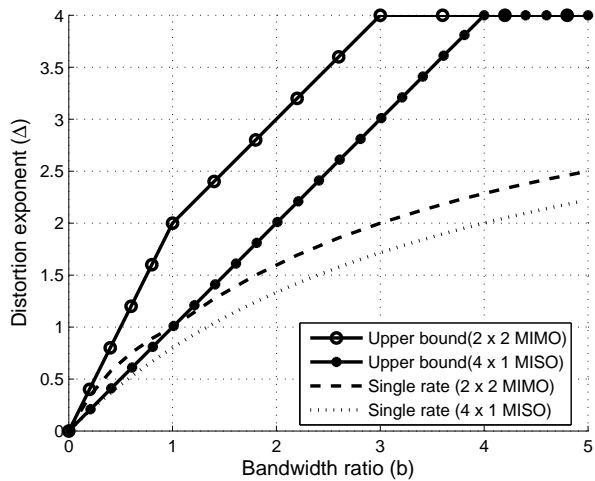}\caption{Upper bound and single layer achievable
distortion exponents for $2 \times 2$ and $4 \times 1$ MIMO
systems.} \label{f:dexp_ubsr}
\end{figure}

In Figure \ref{f:dexp_ubsr}, we illustrate the distortion exponent
upper bound and the distortion exponent of the single layer
scheme for $4\times 1$ MISO and $2\times 2$ MIMO systems. We observe
a significant gap between the upper bounds and the single layer
distortion exponents in both cases for all bandwidth ratios. This gap gets
larger with increasing degrees of freedom and increasing bandwidth ratio.

%---------------------------
%---------------------------

The major drawback of the single layer digital scheme is that it
suffers from the threshold effect, i.e., error probability is
bounded away from zero or an outage occurs when the channel quality
is worse than a certain threshold, which is determined by the
attempted rate. Furthermore, single layer digital transmission
cannot utilize the increase in the channel quality beyond this
threshold. Lack of CSIT makes only a statistical optimization of the
compression/transmission rate possible. To make the system less
sensitive to the variations in the channel quality, we will
concentrate on layered source coding where the channel codewords
corresponding to different layers are assigned different rates.
Using the successive refinability of the source, we transmit more
important compressed bits with higher reliability. The additional
refinement bits are received when the channel quality is high. This
provides adaptation to the channel quality without the transmitter
actually knowing the instantaneous fading levels. We argue that, due
to the exponential decay of the distortion-rate function in general,
layering increases the overall system performance from the
distortion exponent perspective. Our analysis in the following
sections proves this claim.

\section{Layered source coding with progressive transmission and Hybrid Digital-Analog extension}\label{s:LS}

The first source-channel coding scheme we consider is based on
compression of the source in layers, where each layer is a
refinement of the previous ones, and transmission of these layers
successively in time using channel codes of different rates. We call
this scheme \emph{layered source coding with progressive
transmission} (LS). This classical idea, mostly referred as
progressive coding, has been used to various extents in the image
and video standards such as JPEG2000 and MPEG-4. After analyzing the
distortion exponent of LS in Section \ref{ss:LS}, in
Section \ref{ss:HLS} we consider a hybrid digital-analog extension called \emph{hybrid LS} (HLS)
where the error signal is transmitted without coding. In this section we analyze single block fading, i.e., $L=1$, for
clarity of presentation. Generalization to the multiple block case
$(L>1)$ will be a straightforward extension of the techniques
presented here and will be briefly discussed in Section
\ref{s:block_fading}.

%---------------------------
\begin{figure}
\centering
\includegraphics[width=5in]{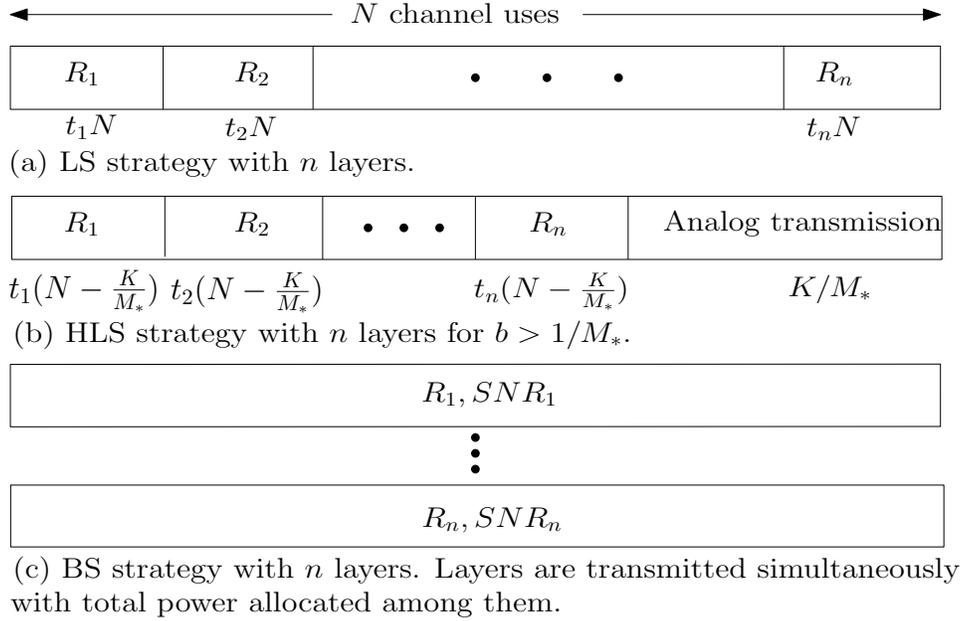}
\caption{Channel and power allocation for different transmission
strategies explored in the paper.} \label{DT2}
\end{figure}
%---------------------------

\subsection {Layered Source Coding with Progressive Transmission (LS)}\label{ss:LS}

We assume that the source encoder has $n$ layers with each layer
transmitted over the channel at rate $R_i$ bits per
channel use in $t_i N$ channel uses for $i=1,2,\ldots,n$, with
$\sum_{i=1}^nt_i=1$. This is illustrated in Fig. \ref{DT2}(a). We assume
that $t_iN$ is large enough to approach the instantaneous channel
capacity. For each layer this corresponds to a source coding rate of
$bt_iR_i$ bits per sample, where $b$ is the bandwidth ratio defined
in (\ref{bexp}). The $i$th layer is composed of the successive
refinement bits for the previous $i-1$ layers. The transmission
power is kept constant for each layer, so the optimization variables
are the rate vector $\mathbf{R}=[R_1, \ldots,R_n]$ and the channel
allocation vector $\mathbf{t}=[t_1,\ldots,t_n]$.

Let $P_{out}^i$ denote the outage probability of layer $i$, i.e.,
$P_{out}^i=Pr\{C(\mathbf{H}) < R_i\}$.  Using successive
refinability of the complex Gaussian source \cite{Equitz}, the
distortion achieved when the first $i$ layers are successfully
decoded is
\begin{eqnarray}
D_i^{LS} &=& D\left(\textit{b}\sum_{k=1}^i
t_kR_k\right), \nonumber \\
&=&2^{-\textit{b}\sum_{k=1}^{i}t_k R_k},
\end{eqnarray}
with $D_0^{LS}=1$. Note that due to  successive refinement source
coding, a layer is useless unless all the preceding layers are
received successfully. This imposes a non-decreasing rate
allocation among the layers, i.e., $R_i\leq R_j$ for any $j> i$.
Then the expected distortion (ED) for such a rate allocation can be
written as
\begin{equation}\label{LS_ED}
ED(\mathbf{R}, \mathbf{t}, SNR) = \sum_{i=0}^{n} D_i^{LS} \cdot
\left(P_{out}^{i+1}- P_{out}^i\right),
\end{equation}
where we define $P_{out}^0=0$ and $P_{out}^{n+1}=1$.

The minimization problem to be solved is
\begin{equation}
\begin{array}{lll}\vspace{.15cm}
\dst{\min_{\mathbf{R}, \mathbf{t}}} & ED(\mathbf{R}, \mathbf{t}, SNR) \\
\vspace{.2cm}
\mbox{s.t.} & \sum_{i=1}^n t_i = 1,\\
\vspace{.2cm}
& t_i\geq0,\mbox{ for } i=1,\ldots,n \\
& 0 \leq R_1\leq R_2\leq \cdots\leq R_n.
\end{array}
\end{equation}
This is a non-linear optimization problem which can be untractable
for a given SNR. An algorithm solving the above optimization
problem for finite SNR is proposed in \cite{Farzad1}. However when
we focus on the high SNR regime and compute the distortion exponent
$\Delta$, we will be able to obtain explicit expressions.

In order to have a vanishing expected distortion in Eqn.
(\ref{LS_ED}) with increasing SNR, we need to increase the
transmission rates of all the layers with SNR as argued in the
single layer case. We let the multiplexing gain vector be
$\mathbf{r}=[r_1,\ldots, r_n]^T$, hence $\mathbf{R}=\mathbf{r}\log
SNR$. The ordering of rates is translated into multiplexing gains as
$0\leq r_1\leq \cdots \leq r_n$. Using the DMT of the MIMO system under consideration and the
distortion-rate function of the complex Gaussian source, we get
\begin{eqnarray}
ED(\mathbf{R},SNR)&\doteq&\sum_{k=0}^n \left[SNR^{-d^*(r_{k+1})}
-SNR^{-d^*(r_k)}\right] SNR^{-\textit{b}\sum_{i=1}^k t_ir_i} \nonumber \\
&\doteq&\sum_{k=0}^n SNR^{-d^*(r_{k+1})} SNR^{-\textit{b}\sum_{i=1}^k t_ir_i} \nonumber \\
&\doteq&SNR^{\max_{0\leq k\leq
n}\left\{-d^*(r_{k+1})-\textit{b}\sum_{i=1}^k t_ir_i\right\},} \label{ED_LS_hSNR}
\end{eqnarray}
where $d^*(r_{n+1})=0$, and the last exponential equality arises
because the summation will be dominated by the slowest decay in high
SNR regime. Then the optimal LS distortion exponent can be written
as
\begin{eqnarray}\label{LS_opt}
\Delta_n^{LS} &=& \max _{\mathbf{r}, \mathbf{t}}\min_{0\leq k\leq
n}\left\{d^*(r_{k+1}) + \textit{b}\sum_{i=1}^k t_i r_i\right\} \\
&\mbox{s.t.}& \sum_{i=1}^n t_i=1, \nonumber \\
&&t_i\geq0\mbox{, for } i=1,\ldots,n \nonumber \\
&& 0\leq r_1\leq r_2\leq \cdots\leq r_n\leq M_*. \nonumber
\end{eqnarray}

Assuming a given channel allocation among $n$ layers, i.e.,
$\mathbf{t}$ is given, the Karush-Kuhn-Tucker (KKT) conditions for
the optimization problem of (\ref{LS_opt}) lead to:
\begin{eqnarray}
b t_n r_n &=& d^*(r_n),  \label{MIMO_st} \\
d^*(r_n)+b t_{n-1} r_{n-1}&=&d^*(r_{n-1}), \label{MIMO_st1} \\
& \dots & \nonumber \\
d^*(r_2)+b t_1 r_1&=&d^*(r_1),\label{MIMO_st2}
\end{eqnarray}
where the corresponding distortion exponent is $\Delta^{LS}_n =
d^*(r_1)$.

%---------------------------
\begin{figure}
\centering
\includegraphics[width=3.5in]{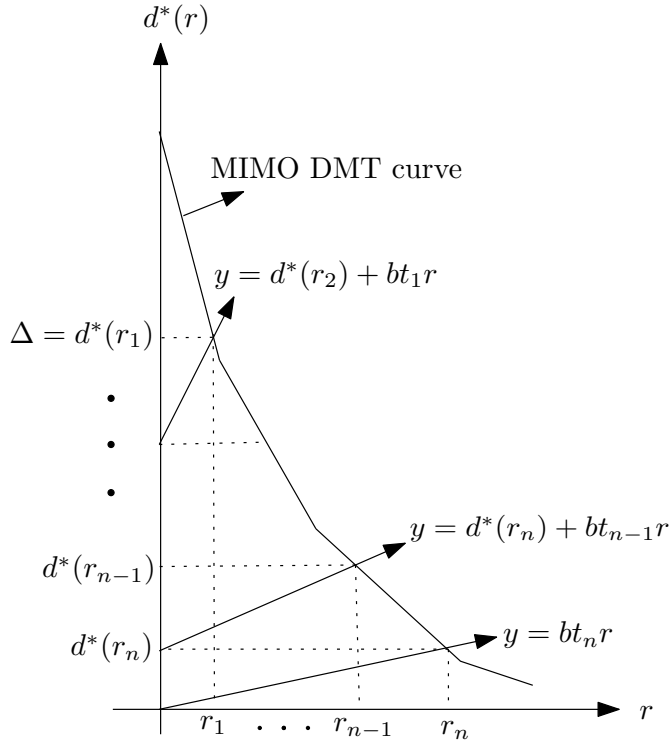}
\caption{Rate allocation for the source layers of LS illustrated on
DMT curve of the MIMO channel.}
\label{MIMO_stair}
\end{figure}
%---------------------------

The equations in (\ref{MIMO_st})-(\ref{MIMO_st2}) can be graphically
illustrated on the DMT curve as shown in
Fig. \ref{MIMO_stair}. This illustration suggests that, for given
channel allocation, finding the distortion exponent in $n$-layer
LS can be formulated geometrically: We have $n$ straight
lines each with slope $\textit{b} t_i$ for $i=1,\ldots,n$, and each
line intersects the $y$-axis at a point with the same ordinate as
the intersection of the previous line with the DMT curve. Although the total slope is
always equal to $b$, the more layers
we have, the higher we can climb on the tradeoff curve and obtain a
larger $\Delta$.

The distortion exponent of LS in the limit of infinite layers
provides a benchmark for the performance of LS in general. The following lemma will be used to characterize the optimal LS
distortion exponent in the limit of infinite layers.

\begin{lem}\label{l:LP_eqall}
In the limit of infinite layers, i.e., as $n \rightarrow \infty$,
the optimal distortion exponent for LS can be achieved by allocating the
channel equally among the layers.
\end{lem}

\begin{proof}
Proof of the lemma can be found in Appendix \ref{A:LP_eqall}.
\end{proof}

The next theorem provides an explicit characterization of the
asymptotic optimal LS distortion exponent $\Delta^{LS}$ (in the case of infinite layers) for an
$M_t\times M_r$ MIMO system.

\begin{thm}\label{delta_LS2}
Let the sequence $\{c_i\}$ be defined as $c_0=0$, $c_i=c_{i-1} +
(|M_r-M_t|+2i-1)\ln \left(\frac{M_*-i+1}{M_*-i}\right)$ for
$i=1,\ldots ,M_*-1$ and $c_{M_*}=\infty$. The optimal distortion exponent of infinite layer
LS is given by:
\begin{eqnarray}\begin{array}{rl}
\Delta^{LS}=\sum_{i=1}^{p-1}&(|M_r-M_t|+2i-1)  \nonumber \\
&+(M_*-p+1)(|M_r-M_t|+2p-1)(1-e^{-\frac{\textit{b}-c_{p-1}}{|M_r-M_t|+2p-1}}),
\end{array}
\end{eqnarray}
for $c_{p-1}\leq
\textit{b}<c_p$, $p=1,\ldots,M_*$.
\end{thm}
\vspace{.01in}
\begin{proof}
Proof of the theorem can be found in Appendix \ref{APP:LS2}.
\end{proof}

\begin{cor}\label{c:delta_LS}
 For a MISO/SIMO system, we have
\begin{equation}\label{dist_exp_LS}
\Delta^{LS}_{MISO/SIMO} = M^*(1-e^{-\mathit{b}/M^*}).
\end{equation}
\end{cor}

Illustration of $\Delta^{LS}$ for some specific examples as well as
comparison with the upper bound and other strategies is left to
Section \ref{s:discussion}. However, we note here that, although LS improves significantly compared to the single layer scheme, it still falls short of the upper bound.  Nevertheless, the advantage of LS is the simple nature of its transceivers. We
only need layered source coding and rate adaptation among layers
while power adaptation is not required.

Another important observation is that, the geometrical model
provided in Fig. \ref{MIMO_stair} and Theorem \ref{delta_LS2} easily
extends to any other system utilizing LS once the DMT
is given. This is done in \cite{SPAWC, relayJ} for a cooperative
system, and will be carried out to extend the results to multiple block fading $(L>1)$ and parallel
channels in Section \ref{s:block_fading}.

\subsection{Hybrid Digital-Analog Transmission with Layered
Source (HLS)}\label{ss:HLS}

In \cite{Caire}, the hybrid digital-analog technique proposed in
\cite{Phamdo} is analyzed in terms of the distortion exponent, and
is shown to be optimal for bandwidth ratios $b\leq 1/M_*$. For
higher bandwidth ratios, while the proposed hybrid strategy
improves the distortion exponent compared to single layer digital
transmission, its performance falls short of the upper bound.
Here, we show that, combining the analog transmission with LS further improves the distortion exponent for $b > 1/M_*$. We call this technique \emph{hybrid digital-analog
transmission with layered source} (HLS). We will show that,
introduction of the analog transmission will improve the distortion
exponent compared to LS with the same number of digital layers,
however; the improvement becomes insignificant as the number of
layers increases.

For $\textit{b}\geq 1/M_*$, we divide the $N$ channel
uses into two portions. In the first portion which is composed of
$N-K/M_*$ channel uses, $n$ source layers are channel coded
and transmitted progressively in time in the same manner as LS. The
remaining $K/M_*$ channel uses are reserved to transmit the error
signal in an analog fashion described below. Channel allocation for
HLS is illustrated in Fig. \ref{DT2}(b).

\begin{figure}
\centering
\includegraphics[width=5in]{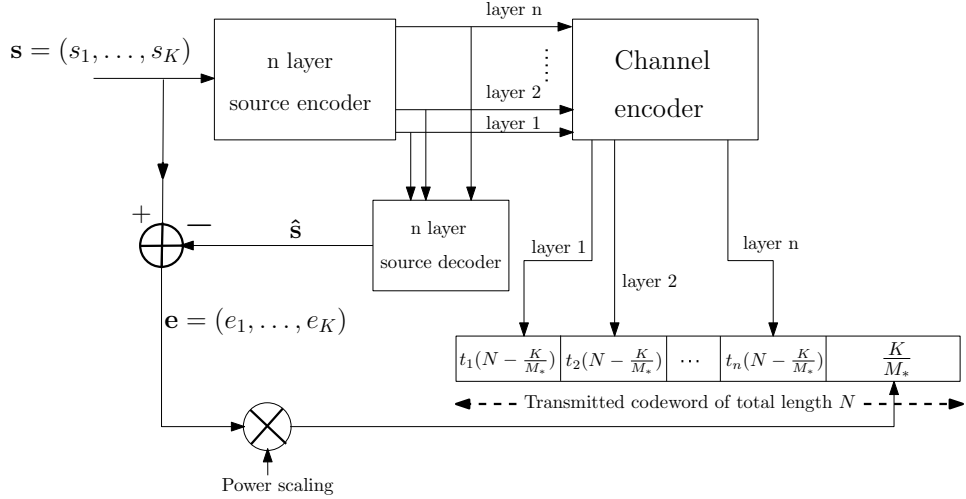}
\caption{Encoder for the $n$ layer HLS for $b>1/M_*$.}
\label{HLS_encoder}
\end{figure}

Let $\bar{\mathbf{s}}\in \mathbb{C}^K$ be the reconstruction of the
source $\mathbf{s}$ upon successful reception of all the digital
layers. We denote the reconstruction error as $\mathbf{e}\in
\mathbb{C}^K$ where $\mathbf{e}=\mathbf{s}-\bar{\mathbf{s}}$. This
error is mapped to the transmit antennas where each component of the
error vector is transmitted without coding in an analog fashion,
just by scaling within the power constraint. Since
$rank(\mathbf{H})\leq M_*$, degrees of freedom of the channel is at
most $M_*$ at each channel use. Hence, at each channel use we
utilize $M_*$ of the $M_t$ transmit antennas and in $K/M_*$ channel
uses we transmit all $K$ components of the error vector
$\mathbf{e}$. HLS encoder is shown in Fig. \ref{HLS_encoder}.

%---------------------------
%---------------------------

Receiver first tries to decode all the digitally transmitted layers as in LS,
and in case of successful reception of all the layers, it forms the
estimate $\bar{\mathbf{s}}+\tilde{\mathbf{e}}$, where
$\tilde{\mathbf{e}}$ is the linear MMSE estimate of $\mathbf{e}$.
This analog portion is ignored unless all digitally transmitted
layers are successfully decoded at the destination.

The expected distortion for $n$-layer HLS can be written
as
\begin{equation}\label{ED_HLS}
ED(\mathbf{R}, SNR) = \sum_{i=0}^{n-1} D_i^{HLS} \cdot
\left(P_{out}^{i+1}-
P_{out}^i\right)+\int_{\mathcal{A}^c}D_a^{HLS}(\bar{\mathbf{H}})
p(\boldsymbol{\lambda}) d\boldsymbol{\lambda},
\end{equation}
where $P_{out}^0=0$, $P_{out}^{n+1}=1$, $D_0^{HLS}=0$,
\begin{equation}
D_i^{HLS}=D\left(\big(\textit{b}-\frac{1}{M_*}\big)\sum_{k=1}^it_kR_k\right)\mbox{
  for } i=1,\ldots,n
\end{equation}
and
\begin{equation}
D_a^{HLS}(\bar{\mathbf{H}})=\frac{D_n^{HLS}}{M_*}\sum_{i=1}^{M_*}\frac{1}{1+\frac{SNR}{M_*}\bar{\lambda}_i},
\end{equation}
where $\mathcal{A}$ denotes the set of channel states at which the
$n$'th layer is in outage,
$\boldsymbol{\lambda}=[\lambda_1,\ldots,\lambda_{M_*}]$ are the
eigenvalues of $\mathbf{H}^\dagger \mathbf{H}$, $\bar{\mathbf{H}}$
is the $M_r \times M_*$ constrained channel matrix, and
$\bar{\boldsymbol{\lambda}}=[\bar{\lambda}_1, \ldots,
\bar{\lambda}_{M_*}]$ are the eigenvalues of
$\bar{\mathbf{H}}^\dagger \bar{\mathbf{H}}$.

Note that the expected distortion of HLS in Eqn. (\ref{ED_HLS})
contains two terms. The first term which consists of the finite sum
can be obtained similar to LS by using appropriate source coding
rates. The following lemma will be used to characterize the high SNR
behavior of the second term.

\begin{lem}\label{l:HLS2}
Suppose that the transmission rate of the $n$-th layer for HLS is $R=r\log SNR$, where $r\leq M_*$ is the multiplexing
gain and that $\mathcal{A}$ denotes the outage event for this layer. If
the average signal-to-noise ratio for the analog part is $SNR$, we have
\begin{eqnarray}
\int_{\mathcal{A}^c}\frac{1}{M_*}\sum_{i=1}^{M_*}\frac{1}{1+
\frac{SNR}{M_*}\bar{\lambda}_i}p(\mathbf{\boldsymbol{\lambda}})
d\mathbf{\boldsymbol{\lambda}} &\dot{\leq}& SNR^{-1}.
\end{eqnarray}
\end{lem}
\vspace{.1in}
\begin{proof}
Proof of the lemma can be found in Appendix \ref{APP:HLS2}.
\end{proof}

Then the second term of the expected distortion in (\ref{ED_HLS})
can be shown to be exponentially less than or equal to
\begin{equation}\label{HLS_decay}
SNR^{-\left[1 + (b-\frac{1}{M_*})\sum_{k=1}^n t_k \right]}.
\end{equation}

Note that at high SNR, both LS and HLS have very similar ED
expressions. Assuming the worst SNR exponent for the second term,
the high SNR approximation of Eqn. (\ref{ED_HLS}) can be written as
in Eqn. (\ref{ED_LS_hSNR}), except for the following: i) the
bandwidth ratio $b$ in (\ref{ED_LS_hSNR}) is replaced by
$b-\frac{1}{M_*}$, and ii) the $n$-th term in (\ref{ED_LS_hSNR}) is
replaced by (\ref{HLS_decay}). Hence, for a given time allocation
vector $\mathbf{t}$, we obtain the following set of equations for
the optimal multiplexing gain allocation:
\begin{eqnarray}
1 + \left(b-\frac{1}{M_*}\right) t_n r_n &=& d^*(r_n),  \label{HLS_st} \\
d^*(r_n)+ \left(b-\frac{1}{M_*}\right) t_{n-1} r_{n-1}&=&d^*(r_{n-1}), \label{HLS_st1} \\
& \dots & \nonumber \\
d^*(r_2)+ \left(b-\frac{1}{M_*}\right) t_1
r_1&=&d^*(r_1),\label{HLS_st2}
\end{eqnarray}
where the corresponding distortion exponent is again $\Delta^{HLS}_n
= d^*(r_1)$. Similar to LS, this formulation enables us to obtain an
explicit formulation of the distortion exponent of infinite layer
HLS using the DMT curve. For brevity we omit the general MIMO HLS distortion exponent and only give the expression for $2 \times 2$ MIMO and general MISO/SIMO systems for comparison.

\begin{cor}
For $2 \times 2$ MIMO, HLS distortion exponent with infinite layers for $b\geq 1/2$ is given by
\begin{eqnarray}
\Delta^{HLS} = 1+3[1-e^{-\frac{1}{3}(b-\frac{1}{2})}]. \nonumber
\end{eqnarray}
\end{cor}
\vspace{.1in}

\begin{cor}
For a MISO/SIMO system utilizing HLS, we have (for $b \geq 1$)
\begin{equation}\label{dist_exp_HLS}
\Delta^{HLS}_{MISO/SIMO} = M^* - (M^*-1) e^{-(b-1)/M^*}.
\end{equation}
\end{cor}

\begin{proof}
For MISO/SIMO, we have $M_*=1$. For $n$
layer HLS with equal time allocation, using Lemma \ref{lemma_LS1} in
Appendix \ref{APP:LS2} we obtain the distortion exponent
\begin{equation}
\hat{\Delta}_{MISO/SIMO,n}^{HLS}=M^*-(M^*-1)\left(\frac{1}{1+\frac{b-1}{nM^*}}\right)^n. \label{eqn5}\\
\end{equation}
Since equal channel allocation in the limit of infinite layers is
optimal, taking the limit as $n \rightarrow \infty$, we obtain (\ref{dist_exp_HLS}).
\end{proof}

Comparing $\Delta^{HLS}_{MISO/SIMO}$ with $\Delta^{LS}_{MISO/SIMO}$
in Corollary \ref{c:delta_LS}, we observe that
\begin{equation}
\Delta^{HLS}_{MISO/SIMO} - \Delta^{LS}_{MISO/SIMO} =
e^{-b/M^*}[M^*-(M^*-1)e^{1/M^*}].
\end{equation}

For a given MISO/SIMO system with a fixed number of $M^*$ antennas,
the improvement of HLS over LS exponentially decays to zero as the
bandwidth ratio increases. Since we have $b \geq 1/M_*$, the biggest
improvement of HLS compared to LS is achieved when $b=1/M_*=1$, and
is equal to $e^{-1/M^*}[M^*-(M^*-1)e^{1/M^*}]$. This is a decreasing
function of $M^*$, and achieves its highest value at $M^*=1$, i.e., SISO system, at $b=1$, and is equal to $1/e$.
Illustration of $\Delta^{HLS}$ for some specific examples as well as
a comparison with the upper bound and other strategies is left to
Section \ref{s:discussion}.

\section{Broadcast Strategy with Layered Source (BS)}\label{s:BS}

In this section we consider superimposing multiple source layers rather than sending them
successively in time. We observe that this leads to higher distortion exponent than LS and HLS, and is in fact optimal in certain cases. This strategy will be called `\emph{broadcast strategy with
layered source}' (BS).

BS combines broadcasting ideas of \cite{Kozintsev1} -\cite{Avi} with layered source coding. Similar to LS, source information is sent in layers,
where each layer consists of the successive refinement information
for the previous layers. As in Section \ref{s:LS} we enumerate the
layers from $1$ to $n$ such that the $i$th layer is the successive
refinement layer for the preceding $i-1$ layers. The codes
corresponding to different layers are superimposed, assigned
different power levels and sent simultaneously throughout the whole transmission
block. Compared to LS, interference among
different layers is traded off for increased multiplexing gain for each layer. We
consider successive decoding at the receiver, where the layers
are decoded in order from $1$ to $n$ and the decoded codewords are
subtracted from the received signal.

Similar to Section \ref{s:LS} we limit our analysis to single block fading $(L=1)$
scenario and leave the discussion of the multiple block case to Section
\ref{s:block_fading}. We first state the
general optimization problem for arbitrary SNR and then study the
high SNR behavior. Let $\mathbf{R} =[R_1, R_2, \ldots, R_n]^T$ be
the vector of channel coding rates, which corresponds to a source
coding rate vector of $\textit{b}\mathbf{R}$ as each code is spread
over the whole $N$ channel uses. Let
$\mathbf{SNR}=[SNR_1,\ldots,SNR_n]^T$ denote the power allocation
vector for these layers with $\sum_{i=1}^n
SNR_i =SNR$. Fig. \ref{DT2}(c) illustrates the channel and power allocation for BS. For $i=1,\ldots,n$ we define
\begin{equation}
\overline{SNR}_i=\sum_{j=i}^n SNR_j.
\end{equation}
The received signal over $N$ channel uses can be written as
\begin{equation}\label{channel_BS}
\mathbf{Y} = \mathbf{H}\sum_{i=1}^n
\sqrt{\frac{SNR_i}{M_t}}\mathbf{X}_i+\mathbf{Z},
\end{equation}
where $\mathbf{Z}\in \mathbb{C}^{M_r\times N}$ is the additive complex
Gaussian noise. We assume each $\mathbf{X}_i\in \mathbb{C}^{M_t\times N}$ is generated
from i.i.d. Gaussian codebooks satisfying
$\textbf{tr}(E[\mathbf{X}_i\mathbf{X}_i^\dag])\leq M_tN$. Here $\sqrt{\frac{SNR_i}{M_t}} \mathbf{X}_i$ carries information for the
$i$-th source coding layer. For $k=1,\ldots,n$, we define
\begin{equation}
\mathbf{\bar{X}}_k = \sum_{j=k}^n \sqrt{\frac{SNR_j}{M_t}}\mathbf{X}_j,
\end{equation}
and
\begin{equation}\label{eq:channel}
\mathbf{Y}_k = \mathbf{H}\mathbf{\bar{X}}_k + \mathbf{Z}.
\end{equation}
Note that $\mathbf{Y}_k$ is the remaining signal at the receiver after decoding and subtracting the first $k-1$
layers. Denoting $\mathcal{I}(\mathbf{Y}_k;\mathbf{X}_k)$ as the mutual information between $\mathbf{Y}_k$ and $\mathbf{X}_k$, we can define the following outage events,
\begin{eqnarray}
\mathcal{A}_k &=& \left\{\mathbf{H}:\mathcal{I}(\mathbf{Y}_k;\mathbf{X}_k) < R_k\right\}, \label{d:A_k}\\
\mathcal{B}_k &=& \bigcup_{i=1}^k \mathcal{A}_i, \label{d:B_k}
\end{eqnarray}
and the corresponding outage probabilities
\begin{eqnarray}
P_{out}^k &=& Pr\left\{\mathbf{H}:\mathbf{H}\in \mathcal{A}_k \right\},\\
\bar{P}_{out}^k &=& Pr\left\{\mathbf{H}:\mathbf{H}\in \mathcal{B}_k
\right\}. \label{pout_bar}
\end{eqnarray}
We note that $P_{out}^k$ denotes the probability of outage for layer
$k$ given that the decoder already has access to the previous $k-1$ layers. On the other hand,
$\bar{P}_{out}^k$ is the overall outage probability of layer $k$ in
case of successive decoding, where we assume that
if layer $k$ cannot be decoded, then the receiver will not attempt to decode the subsequent layers $i>k$. Then the expected distortion for
$n$-layer BS using successive decoding  can be written as
\begin{eqnarray}\label{BS_ED}
ED(\mathbf{R}, \mathbf{SNR})&=& \sum_{i=0}^n
D_i^{BS}(\bar{P}_{out}^{i+1} -\bar{P}_{out}^i ) ,
\end{eqnarray}
where \[D_i^{BS} = D\left(\textit{b}\sum_{k=1}^i R_k\right),\]
$\bar{P}_{out}^0 = 0$, $\bar{P}_{out}^{n+1} = 1$, and $D_0^{BS}=1$. Various algorithms solving this optimization problem are proposed in \cite{Honig},\cite{Farzad2}-\cite{Tian}.

The following definition will be useful in characterizing the distortion exponent of BS.

\begin{defn}
The `\emph{successive decoding diversity gain}' for layer $k$ of the BS strategy is defined as the high SNR exponent of the outage probability of that layer using successive decoding at the receiver. The successive decoding diversity gain can be written as
\[d_{sd}(r_k) \triangleq - \lim_{SNR\rightarrow \infty} \frac{\log \bar{P}_{out}^{k}}{\log SNR}\]
\end{defn}
\vspace{.1in}
Note that the successive decoding diversity gain for layer $k$ depends on the power and multiplexing gain allocation for layers $1,\ldots,k-1$ as well as layer $k$ itself. However, we drop the dependence on the previous layers for simplicity.

For any communication system with DMT characterized by $d^*(r)$, the
successive decoding diversity gain for layer $k$ satisfies
$d_{sd}(r_k) \leq d^*(r_1+\ldots + r_k)$. Concurrent work by Diggavi
and Tse \cite{Diggavi1}, coins the term `\emph{successive
refinability of the DMT curve}' when this inequality is satisfied
with equality, i.e. multiple layers of information simultaneously
operate on the DMT curve of the system. Our work, carried out
independently, illustrates that combining successive refinability of
the source and the successive refinability of the
DMT curve leads to an optimal distortion
exponent in certain cases.

From (\ref{BS_ED}), we can write the high SNR approximation for $ED$ as below.
\begin{eqnarray}
ED &\doteq& \sum_{i=0}^n \bar{P}_{out}^{i+1} D_i^{BS}, \nonumber \\
&\doteq& \sum_{i=0}^n SNR^{-d_{sd}(r_{i+1})} SNR^{-b\sum_{j=1}^i
r_j}. \label{BS_ED_hSNR}
\end{eqnarray}
Then the distortion exponent is given by
\begin{eqnarray}\label{BS_ED_de}
\Delta_n^{BS} = \min_{0\leq i \leq n} \left\{d_{sd}(r_{i+1})+ b\sum_{j=1}^i r_j \right\}.
\end{eqnarray}
Note that, while the DMT curve for a given system is enough to find the corresponding distortion exponent for LS and HLS, in the case of BS, we need the successive decoding DMT curve.

Next, we propose a power allocation among layers for a given multiplexing gain vector. For a general $M_t \times M_r$ MIMO system, we consider
multiplexing gain vectors $\mathbf{r}=[r_1,\ldots, r_n]$such that $r_1+\cdots+r_n \leq 1$. This constraint ensures that we obtain an
increasing and nonzero sequence of outage probabilities
$\{P_{out}^k\}_{k=1}^n$. We impose the following power allocation among the layers:
\begin{eqnarray}\label{pow_alloc_MIMO}
\overline{SNR}_k=SNR^{1-(r_1+\cdots+r_{k-1}+\epsilon_{k-1})},
\end{eqnarray}
for $k=2,\ldots,n$ and $0<\epsilon_1<\cdots<\epsilon_{n-1}$.

Our next theorem computes the successive decoding diversity gain obtained with the above power allocation. We will see that the proposed power allocation scheme results in successive refinement of the DMT curve for MISO/SIMO systems. By optimizing the multiplexing gain $\mathbf{r}$ we will show that the optimal distortion exponent for MISO/SIMO meets the distortion exponent upper bound.

\begin{thm}\label{t:MIMO_DMT}
For $M_t \times M_r$ MIMO, the successive decoding diversity gain for the power allocation in (\ref{pow_alloc_MIMO}) is given by
\begin{eqnarray}
d_{sd}(r_k) = M^*M_*(1-r_1-\cdots-r_{k-1}) - (M^*+M_*-1)r_k.
\end{eqnarray}
\end{thm}

\begin{proof}
Proof of the theorem can be found in Appendix \ref{APP:MIMO_DMT}.
\end{proof}

\begin{cor}\label{c:MISO}
The power allocation in (\ref{pow_alloc_MIMO}) results in the successive refinement of the DMT curve for MISO/SIMO systems.
\end{cor}

\begin{proof}
For MISO/SIMO we have $M_*=1$. By Theorem \ref{t:MIMO_DMT} we have $d_{sd}(r_k) = M^* (1-r_1-\cdots-r_k) = d^*(r_1+\cdots +r_k)$. Thus all simultaneously transmitted $n$ layers operate on the DMT curve.
\end{proof}

Using the successive decoding DMT curve of Theorem \ref{t:MIMO_DMT}, the next theorem computes an achievable distortion exponent for BS by optimizing the multiplexing gain allocation among layers.

\begin{thm}\label{t:MIMO_DE}
For $M_t \times M_r$ MIMO, $n$-layer BS with power allocation in (\ref{pow_alloc_MIMO}) achieves a distortion exponent of
\begin{eqnarray}
\Delta^{BS}_n = b\frac{(M_t-k)(M_r-k)(1-\eta_k^n)}{(M_t-k)(M_r-k)-\textit{b}\eta_k^n},
\end{eqnarray}
for $b \in \big[(M_t-k-1)(M_r-k-1), (M_t-k)(M_r-k)\big)$,
$k=0,\ldots,M_*-1$, where
\begin{eqnarray}
\eta_k = 1 + \frac{b-(M_t-k-1)(M_r-k-1)}{M_t + M_r-2k-1} >0,
\label{eta_k}
\end{eqnarray}
and
\begin{eqnarray}
\Delta^{BS}_n =  \frac{n(M_tM_r)^2}{nM_tM_r+M_t+M_r-1},
\end{eqnarray}
for $b \geq M_tM_r$.

In the limit of infinite layers, BS distortion exponent becomes
\begin{eqnarray}
\Delta^{BS} = \left\{
\begin{array}{lll}
                \textit{b} & \mbox{if} & \textit{b} < M_tM_r, \vspace{.02in} \\
                M_tM_r & \mbox{if} & b \geq M_tM_r. \\
           \end{array} \right.
\end{eqnarray}
Hence, BS is distortion exponent optimal for $b \geq M_tM_r$.
\end{thm}
\vspace{.1in}
\begin{proof}
Proof of the theorem can be found in Appendix \ref{APP:MIMO_DE}.
\end{proof}

\begin{cor}\label{c:BS1}
For a MISO/SIMO system, the $n$-layer BS distortion exponent achieved by the power allocation in (\ref{pow_alloc_MIMO}) is
\begin{equation}\label{eqn12}
\Delta^{BS}_{MISO/SIMO,n} =M^*\left(1-\frac{1-\mathit{b}/M^*}{1-(\mathit{b}/M^*)^{n+1}}\right).
\end{equation}\vspace{.01in}
In the limit of infinite layers, we obtain
\begin{equation}\label{dist_exp_MISO}
\Delta^{BS}_{MISO/SIMO} = \left\{ \begin{array}{lll}
            \mathit{b} & \mbox{if} & \mathit{b}< M^*, \vspace{.02in} \\
              M^* & \mbox{if} & \mathit{b} \geq M^*. \\
           \end{array} \right.
\end{equation}
BS with infinite source layers meets the distortion exponent upper bound of MISO/SIMO given in (\ref{UB_MISO}) for all bandwidth ratios, hence is optimal. Thus, (\ref{dist_exp_MISO}) fully characterizes the optimal distortion exponent for MISO/SIMO systems.
\end{cor}
\vspace{.1in}

Recently, \cite{Bhattad1, Bhattad2} reported improved BS distortion exponents for general MIMO by a more advanced power allocation strategy. Also, while a successively refinable DMT would increase the distortion exponent, we do not know whether it is essential to achieve the distortion exponent upper bound given in Theorem \ref{t:upperbound}.  However, successive refinement of general MIMO DMT has not been established \cite{Diggavi1, Diggavi2}.

As in Section \ref{s:LS}, the discussion of the results is left to Section \ref{s:discussion}.

%---------------------------
\begin{figure}
\centering
\includegraphics[width=3.5in]{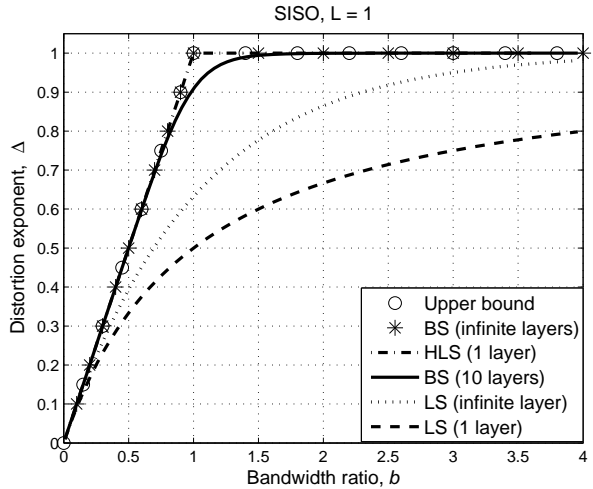}
\caption{Distortion exponent vs. bandwidth ratio for SISO channel,
$L=1$.} \label{SISO}
\end{figure}
%---------------------------

\section{Multiple Block Fading Channel} \label{s:block_fading}

In this section, we extend the results for $L=1$ to multiple block
fading MIMO, i.e., $L>1$. As we observed throughout the previous
sections, the distortion exponent of the system is strongly related
to the maximum diversity gain available over the channel. In the
multiple block fading scenario, channel experiences $L$ independent
fading realizations during $N$ channel uses, so we have $L$ times
more diversity as reflected in the DMT of Theorem \ref{divmimo}. The
distortion exponent upper bound in Theorem \ref{t:upperbound}
promises a similar improvement in the distortion exponent for
multiple block fading. However, we note that increasing $L$ improves
the upper bound only if the bandwidth ratio is greater than
$|M_t-M_r|+1$, since the upper bound is limited by the bandwidth
ratio, not the diversity at low bandwidth ratios.

Following the discussion in Section \ref{s:LS}, extension of LS
to multiple block fading is straightforward. As before, each layer
of the successive refinement source code should operate on a
different point of the DMT. This
requires us to transmit codewords that span all channel
realizations, thus we divide each fading block among $n$ layers and transmit the codeword of each layer over its $L$ portions.

%---------------------------
\begin{figure}
\centering
\includegraphics[width=3.5in]{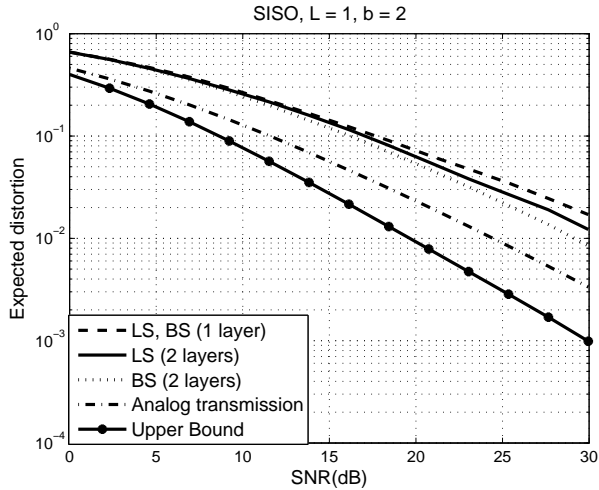}
\caption{Expected distortion vs. SNR plots for $\mathit{b}=2$. The
topmost curve LS, BS (1 layer) corresponds to single layer
transmission.} \label{ED_SISO}
\end{figure}
%---------------------------

Without going into the details of an explicit derivation of the
distortion exponent for the multiple block
fading case, as an example, we find the LS distortion exponent for $2\times 2$ MIMO
channel with $L=2$ as
\begin{eqnarray}
\Delta^{LS} = \left\{
\begin{array}{lll}
               4(1-\exp(-\textit{b}/2)) & \mbox{if} & 0<\textit{b}\leq 2\ln2, \vspace{.02in} \\
              2+6\left[1-\exp\left(-(\frac{\textit{b}}{6}-\frac{\ln2}{3})\right)\right] & \mbox{if} & \textit{b}> 2\ln2. \\
           \end{array} \right.
\end{eqnarray}

For HLS, the threshold
bandwidth ratio $1/M_*$ is the same as single block fading
as it depends on the channel rank per channel use, not the
number of different channel realizations. For $\textit{b}\geq1/M_*$,
extension of HLS to multiple blocks can be done similar to LS. As an example, for 2-block Rayleigh fading $2\times 2$ MIMO channel, the optimal distortion
exponent of HLS for $\textit{b}\geq1/2$ is given by
\begin{eqnarray}
\Delta^{HLS} = \left\{
\begin{array}{lll}
               1 + 3\left[1-\exp\left(-\frac{1}{2}(\textit{b}-\frac{1}{2})\right)\right] & \mbox{if} & 1/2\leq\textit{b}\leq \frac{1}{2} + 2\ln\frac{3}{2}, \vspace{.02in} \\
              2 + 6\left[1-\exp\left(-\frac{1}{6}(\textit{b}-\frac{1}{2}-2\ln\frac{3}{2})\right)\right] & \mbox{if} & \textit{b}> \frac{1}{2}+2\ln\frac{3}{2}. \\
           \end{array} \right.
\end{eqnarray}

For BS over multiple block fading MIMO, we
use a generalization of the power allocation introduced in Section
\ref{s:BS} in (\ref{pow_alloc_MIMO}). For $L$-block fading channel and for $k=2,\ldots,n$ let
\begin{eqnarray}\label{pow_alloc_BF}
\overline{SNR}_k = SNR^{1-L(r_1+\cdots+r_{k-1}-\epsilon_{k-1})},
\end{eqnarray}
with $0<\epsilon_1<\cdots <\epsilon_{n-1}$ and imposing $\sum_{i=1}^n r_i \leq 1/L$. Using this power allocation scheme, we obtain the following distortion exponent for BS over $L$-block MIMO channel.

\begin{thm}\label{t:block_DE}
For $L$-block $M_t \times M_r$ MIMO, BS with power allocation in (\ref{pow_alloc_BF}) achieves the following distortion exponent in the limit of infinite layers.
\begin{eqnarray}
\Delta^{BS} = \left\{
\begin{array}{lll}
                b/L & \mbox{if} & \textit{b} < L^2M_tM_r, \vspace{.02in} \\
                LM_tM_r & \mbox{if} & b \geq L^2M_tM_r. \\
           \end{array} \right.
\end{eqnarray}
This distortion exponent meets the upper bound for $b \geq L^2M_tM_r$.
\end{thm}

\begin{proof}
Proof of the theorem can be found in Appendix \ref{APP:block_DE}.
\end{proof}

The above generalizations to multiple block
fading  can be adapted to parallel channels
through a scaling of the bandwidth ratio by $L$ \cite{ISIT06}. Note that, in the
block fading model, each fading block
lasts for $N/L$ channel uses. However, for $L$ parallel channels with independent fading, each block lasts
for $N$ channel uses instead. Using the power allocation in (\ref{pow_alloc_MIMO}) we can get achievable BS distortion exponent for parallel channels. We refer the reader to \cite{ISIT06} for details and comparison. Detailed discussion and comparison of the $L$-block LS, HLS and BS distortion exponents are left to Section \ref{s:discussion}.

Distortion exponent for parallel channels has also been studied in \cite{Emin} and \cite{Dunn} both of which consider source and channel coding
for two parallel fading channels. The analysis of \cite{Emin} is
limited to single layer source coding and multiple description
source coding. Both schemes perform worse than the upper
bound in Theorem \ref{t:upperbound} and the achievable strategies presented in this paper. Particularly, the best distortion exponent achieved in \cite{Emin} is by single layer source coding
and parallel channel coding which is equivalent to LS
with one layer. In \cite{Dunn}, although 2-layer successive
refinement and hybrid digital-analog transmission are considered,
parallel channel coding is not used, thus the achievable performance is limited.
The hybrid scheme proposed in \cite{Dunn} is a repetition based
scheme and cannot improve the distortion exponent beyond single layer LS.

%---------------------------
\begin{figure}
\centering
\includegraphics[width=3.5 in]{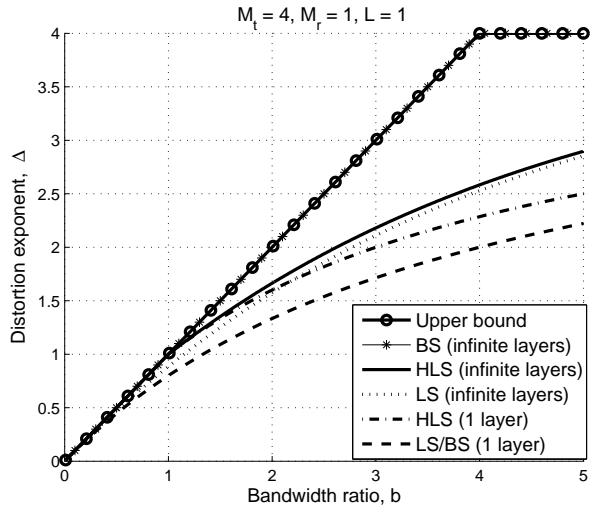}
\caption{Distortion exponent vs. bandwidth ratio for $4\times 1$
MIMO, $L=1$.} \label{MIMO4x1}
\end{figure}
%---------------------------
\section{Discussion of the Results}\label{s:discussion}

This section contains a discussion and comparison of all the schemes
proposed in this paper and the upper bound. We first consider the
special case of single-input single-output (SISO) system. For a SISO
single block Rayleigh fading channel, the upper bound for optimal
distortion exponent in Theorem \ref{t:upperbound} can be written as
\begin{equation}
\Delta= \left\{ \begin{array}{lll}
              \mathit{b}  & \mbox{if} & \textit{b}<1, \vspace{.02in} \\
              1 & \mbox{if} & \textit{b}\geq 1. \\
           \end{array} \right.
\end{equation}
This optimal distortion exponent is achieved by BS in the limit of
infinite source coding layers (Corollary \ref{c:MISO}) and by HLS
\cite{Caire}. In HLS, pure analog transmission is enough to reach
the upper bound when $\textit{b}\geq 1$ \cite{Asil}, while the
hybrid scheme of \cite{Caire} achieves the optimal distortion
exponent for $\textit{b}<1$.

The distortion exponent vs. bandwidth ratio of the various schemes
for the SISO channel, $L=1$ are plotted in Fig. \ref{SISO}. The
figure suggests that while BS is optimal in the limit of infinite
source layers, even with 10 layers, the performance is very close to
optimal for almost all bandwidth ratios.

For a SISO channel, when the performance measure is the expected
channel rate,  most of the improvement provided by the broadcast
strategy  can be obtained with two layers \cite{Fitz}. However, our
results show that when the performance measure is the expected
end-to-end distortion, increasing the number of superimposed layers
in BS further improves the performance especially for bandwidth
ratios close to 1.
%---------------------------
\begin{figure}
\centering
\includegraphics[width=3.5in]{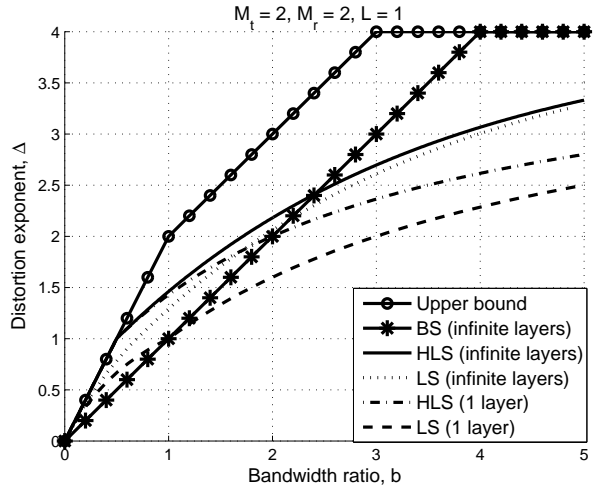}
\caption{Distortion exponent vs. bandwidth ratio for $2\times 2$
MIMO, $L=1$.} \label{MIMO2x2}
\end{figure}
%---------------------------

In order to illustrate how the suggested source-channel coding
techniques perform for arbitrary $SNR$ values for the SISO channel,
in Fig. \ref{ED_SISO} we plot the expected distortion vs. $SNR$ for
single layer transmission (LS, BS with 1 layer), LS and BS with $2$
layers, analog transmission and the upper bound for $\mathit{b}=2$.
The results are obtained from an exhaustive search over all possible
rate, channel and power allocations. The figure illustrates that the
theoretical distortion exponent values that were found as a result
of the high $SNR$ analysis hold, in general, even for moderate $SNR$
values.

In Fig. \ref{MIMO4x1}, we plot the distortion exponent versus
bandwidth ratio of $4\times 1$ MIMO single block fading channel for
different source-channel strategies discussed in Section
\ref{s:LS}-\ref{s:BS} as well as the upper bound. As stated in
Corollary \ref{c:MISO}, the distortion exponent of BS coincides with
the upper bound for all bandwidth ratios. We observe that HLS is
optimal up to a bandwidth ratio of $1$. This is attractive for
practical applications since only a single coded layer is used,
while BS requires many more layers to be superimposed. However, the
performance of HLS degrades significantly beyond $b=1$, making BS
more advantageous in this region. Pure analog transmission of the
source samples would still be limited to a distortion exponent of
$1$ as in SISO, since linear encoding/decoding can not
utilize the diversity gain of the system. More advanced nonlinear
analog schemes which would take advantage of the diversity gain and
achieve an improved distortion exponent may be worth exploring.
While LS does not require any superposition or power allocation
among layers, and only uses a digital encoder/decoder pair which can
transmit at variable rates, the performance is far below BS.
Nevertheless, the improvement of infinite layer LS compared to a
single layer strategy is significant.

%---------------------------
\begin{figure}
\centering
\includegraphics[width=3.5in]{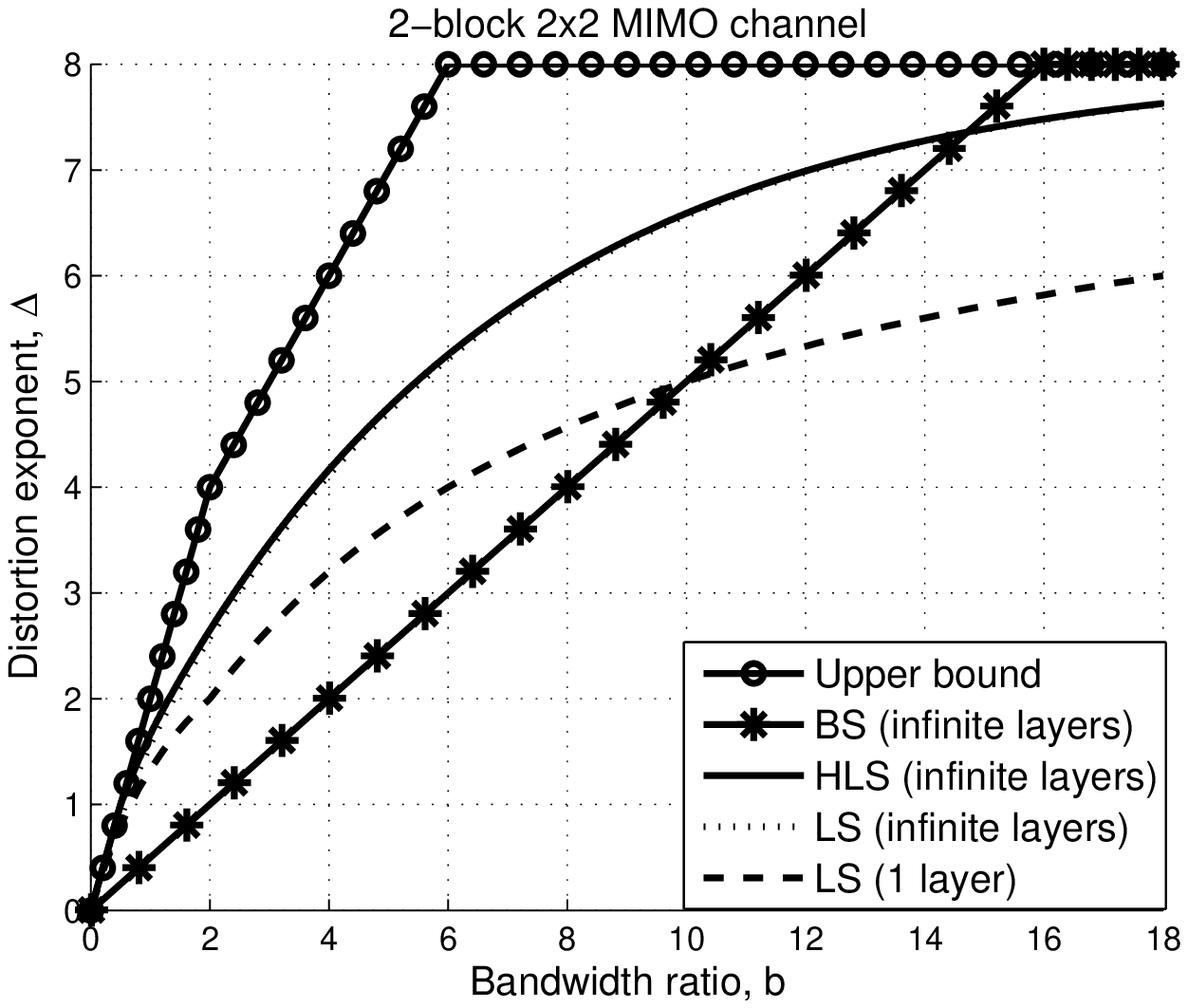}
\caption{Distortion exponent vs. bandwidth ratio for 2x2 MIMO,
$L=2$.} \label{2block}
\end{figure}
%---------------------------

We plot the distortion exponent versus bandwidth ratio for $2\times
2$ MIMO with $L=1$ in Fig. \ref{MIMO2x2}. We observe that BS is
optimal for $\mathit{b}\geq 4$ and provides the best distortion
exponent for $b>2.4$. For bandwidth ratios $1/2<\mathit{b}< 4$, none
of the strategies discussed in this paper achieves the upper bound.

Note that, both for $4\times 1$ MISO and $2\times 2$ MIMO, when
$\mathit{b}\geq 1/M^*$ the gain due to the analog portion, i.e.,
gain of HLS compared to LS, is more significant for one
layer and decreases as the number of layers goes to infinity.
Furthermore, at any fixed number of layers, this gain decays to zero
with increasing bandwidth ratio as well. We conclude that for
general MIMO systems, when the bandwidth ratio is high, layered
digital transmission with large number of layers results in the
largest improvement in the distortion exponent.

In Figure \ref{2block} we plot the distortion exponent for a
$2$-block $2\times2$ MIMO channel. We observe that the improvement
of HLS over LS, both operating with infinite number of layers, is even
less significant than the single block case. However, HLS can still achieve the
optimal distortion exponent for $\mathit{b}<1/2$. Although BS is
optimal for $\mathit{b}\geq L^2M_tM_r=16$, both this threshold of
$L^2M_tM_r$ and the gap between the upper bound and BS performance
below this threshold increases as $L$ increases.

\section{Generalization to Other Sources}\label{sectgen}

Throughout this paper, we have used a complex Gaussian source for
clarity of the presentation. This assumption enabled us to use the
known distortion-rate function and to utilize the successive
refinable nature of the complex Gaussian source. In this section we
argue that our results hold for any memoryless source with finite
differential entropy and finite second moment under squared-error
distortion.

Although it is hard to explicitly obtain the rate-distortion
function of general stationary sources, lower and upper bounds
exist. Under the mean-square error distortion criteria, the
rate-distortion function $R(D)$ of any stationary continuous
amplitude source $X$ is bounded as \cite{Berger}
\begin{eqnarray}
R_L(D) \leq R(D) \leq R_G(D),
\end{eqnarray}
where $R_L(D)$ is the Shannon lower bound and $R_G(D)$ is the rate-distortion function of the complex
Gaussian source with the same real and imaginary variances.

Further in \cite{Zamir} it is shown that the Shannon lower bound is
tight in the low distortion $(D\rightarrow 0)$, or, the high rate
$(R\rightarrow\infty)$ limit when the source has a finite moment and
finite differential entropy. We have
\begin{eqnarray}
\lim_{R\rightarrow\infty}D(R)-\frac{e^{2h(X)}}{2\pi e}2^{-R}=0.
\end{eqnarray}

%In one of the important results of the high resolution quantization
%theory \cite{Zador}, Zador considers variable-rate scalar
%quantization in the high rate limit, and finds the distortion-rate
%function as
%\begin{eqnarray}
%D(R)\sim \frac{1}{12}2^{2h(X)-2R}.
%\end{eqnarray}

The high rate approximation of the distortion-rate function can be
written as $D(R)=2^{-R+O(1)}$, where $O(1)$ term depends on the
source distribution but otherwise independent of the compression
rate. Since in our distortion exponent analysis we consider scaling
of the transmission rate, hence the source coding rate,
logarithmically with increasing SNR, the high resolution
approximations are valid for our investigation. Furthermore the
$O(1)$ terms in the above distortion-rate functions do not change
our results since we are only interested in the SNR exponent of the
distortion-rate function.

Although most sources are not successively refinable, it was proven
in \cite{Lastras} that all sources are nearly successively
refinable. Consider $n$ layer source coding with rate of $R_i$
bits/sample for layer $i=1,\ldots,n$. Define $D_i$ as the distortion
achieved with the knowledge of first $i$ layers and $W_i=R_i-R(D_i)$
as the rate loss at step $i$, where $R(D)$ is the distortion-rate
function of the given source. Throughout the paper we used the fact
that this rate loss is $0$ for the Gaussian source \cite{Equitz}.
Now we state the following result from \cite{Lastras} to argue that
our high SNR results hold for sources that are nearly successively
refinable as well.

\begin{lem}(Corollary 1, \cite{Lastras})
For any $0<D_n<...<D_2<D_1$, $(n\geq2)$ and squared error
distortion, there exists an achievable M-tuple $(R_1,\ldots,R_n)$
with $W_k\leq1/2$, $k\in\{1,...,n\}$.\\
\end{lem}

This means that to achieve the distortion levels we used in our
analysis corresponding to each successive refinement layer, we need
to compress the source layer at a rate that is at most
$1$ bits/sample\footnote{We have $W_k\leq1$ due to complex source assumption.} greater than the rates required for a successively refinable source. This translates into the distortion rate function
as an additional $O(1)$ term in the exponent, which, as argued
above, does not change the distortion exponent results obtained in
this paper. These arguments together suggest that relaxing the
Gaussian source assumption alters neither the calculations nor the
results of our paper.

\section{Conclusion}\label{s:conclusion}

We considered the fundamental problem of joint source-channel coding
over block fading MIMO channels in the high SNR regime with no CSIT
and perfect CSIR. Although the general problem of characterizing the achievable average
distortion for finite SNR is still open, we showed that we
can completely specify the high SNR behavior of the expected
distortion in various settings. Defining the distortion exponent as
the decay rate of average distortion with SNR, we provided a
distortion exponent upper bound and three different lower bounds.

Our results reveal that, layered source coding with unequal error protection is critical for adapting to the variable channel state without the availability of CSIT. For the proposed transmission schemes, depending on the bandwidth ratio, either progressive or simultaneous transmission of the layers perform better. However, for single block MISO/SIMO channels, BS outperforms all other strategies and meets the upper bound, that is, BS is distortion exponent optimal for MISO/SIMO.

\appendices
\section{Proof of Theorem \ref{t:upperbound}  } \label{APP:UB}

Here we find the distortion exponent upper bound under short-term
power constraint assuming availability of the channel state
information at the transmitter (CSIT)\footnote{We note that a
similar upper bound for $L=1$ is also given in \cite{Caire}. We
derive it for the $L$-block channel here for completeness.}. Let
$C(\mathbf{H})$ denote the capacity of the channel with short-term
power constraint when CSIT is present. Note that $C(\mathbf{H})$
depends on the channel realizations
$\mathbf{H}_1,\ldots,\mathbf{H}_L$. The capacity achieving input
distribution at channel realization $\mathbf{H}_j$ is Gaussian with
covariance matrix $\mathbf{Q}_j$. We have
\begin{eqnarray}
C(\mathbf{H})&=&\frac{1}{L}\sum_{j=1}^L\sup_{\mathbf{Q}_j\succeq 0,
\sum_{j=1}^L\textbf{tr}(\mathbf{Q}_j)\leq LM_t} \log \det
\left(\textbf{I}+\frac{SNR}{M_t}\mathbf{H}_j\mathbf{Q}_j\mathbf{H}_j^\dagger\right),
\nonumber\\
&\leq&\frac{1}{L}\sum_{j=1}^L\sup_{\mathbf{Q}_j\succeq 0,
\textbf{tr}(\mathbf{Q}_j)\leq LM_t} \log \det
\left(\textbf{I}+\frac{SNR}{M_t}\mathbf{H}_j\mathbf{Q}_j\mathbf{H}_j^\dagger\right),
\nonumber\\
&\leq& \frac{1}{L}\sum_{j=1}^L\log\det(\textbf{I}+L\cdot
SNR\mathbf{H}_j\mathbf{H}_j^\dagger),
\end{eqnarray}
where the first inequality follows as we expand the search space,
and the second inequality follows from the fact that
$LM_t\textbf{I}-\mathbf{Q}_j\succeq0$ when
$\textbf{tr}(\mathbf{Q}_j)\leq LM_t$ and $\log\det(\cdot)$ is an
increasing function on the cone of positive-definite Hermitian
matrices. Then the end-to-end distortion can be lower bounded as
\begin{equation}\label{eqnDH}
D(\mathbf{H})=2^{-bC(\mathbf{H})} \geq\prod_{j=1}^L
[\det(\textbf{I}+L\cdot
SNR\mathbf{H}_j\mathbf{H}_j^\dagger)]^{-\textit{b}/L}.
\end{equation}

We consider expected distortion, where the expectation is taken over
all channel realizations and analyze its high SNR exponent to find
the corresponding distortion exponent. We will follow the technique
used in \cite{Tse}. Assume without loss of generality that $M_t\geq
M_r$. Then from Eqn. (\ref{eqnDH}) we have
\begin{eqnarray}
D(\mathbf{H}) &\geq& \prod_{j=1}^{L}[\det (\mathbf{I} + L\cdot SNR
  \mathbf{H}_j\mathbf{H}_j^\dag)]^{-\textit{b}/L}, \\
       &\geq& \prod_{j=1}^{L} \prod_{i=1}^{M_r}(1+L\cdot SNR\lambda_{ji})^{-\textit{b}/L},
\end{eqnarray}
where $\lambda_{j1} \leq\lambda_{j1} \leq \dots \leq\lambda_{jM_r}$
are the ordered eigenvalues of $\mathbf{H}_j\mathbf{H}_j^\dag$ for
block $j=1,\ldots,L$. Let $\lambda_{ji}=SNR^{-\alpha_{ji}}$. Then we
have $(1+ L \cdot SNR\lambda_{ji})\doteq SNR^{(1-\alpha_{ji})^+}$.

The joint pdf of $\boldsymbol{\alpha_j}=[\alpha_{j1}, \ldots,
\alpha_{jM_t}]$ for $j=1,\ldots,L$ is
\begin{eqnarray}
p(\boldsymbol{\alpha}_j)&=& K_{M_t, M_r}^{-1}(\log SNR)^{M_r}\prod
_{i=1}^{M_r}SNR^{-(M_t-M_r+1)\alpha_{ji}} \nonumber\\
&& \cdot\left[
\prod_{i<k}(SNR^{\alpha_{ji}}-SNR^{\alpha_{jk}})^2\right]\exp
\left(-\sum_{i=1}^{M_r}SNR^{\alpha_{ji}}\right),
\end{eqnarray}
where $K_{M_t, M_r}$ is a normalizing constant. We can write the
expected end-to-end distortion as
\begin{eqnarray}
E[D(\mathbf{H})] &\doteq& \int D(\mathbf{H})
p(\boldsymbol{\alpha}_1)\ldots
p(\boldsymbol{\alpha}_L)d\boldsymbol{\alpha}_1\ldots d\boldsymbol{\alpha}_L,\\
 &\dot{\geq}& \left[\int
\prod_{i=1}^{M_r}(1+SNR\lambda_{1i})^{-\textit{b}/L}p(\boldsymbol{\alpha}_1)d\boldsymbol{\alpha}_1\right]^L,
\end{eqnarray}
where we used the fact that $\boldsymbol{\alpha}_j$'s are i.i.d. for
$j=1,\ldots,L$ and that the constant multiplicative term in front of
SNR does not affect the exponential behavior. Since we are
interested in the distortion exponent, we only need to consider the
exponents of SNR terms. Following the same arguments as in
\cite{Tse} we can make the following simplifications.
\begin{equation}
\begin{array}{lcl}
\dst{\int\prod_{i=1}^{M_r}(1+SNR\lambda_{1i})^{-\textit{b}/L}p(\boldsymbol{\alpha}_1)d\boldsymbol{\alpha}_1}
 &\doteq&\dst{ \int_{\mathcal{R}^{n+}}\prod_{i=1}^{M_r}(1+SNR^{1-\alpha_{1i}})^{-\textit{b}/L}}\nonumber \\
\end{array}
\end{equation}
\begin{equation}
\begin{array}{ll}
&\dst{\cdot \prod_{i=1}^{M_r}SNR^{-(M_t-M_r+1) \alpha_{1i}} \cdot\prod_{i<k}(SNR^{-\alpha_{1i}}-SNR^{-\alpha_{1k}})^2 d\boldsymbol{\alpha}_1.} \nonumber \\
 \end{array}
 \end{equation}
\begin{eqnarray}
&\doteq&\int_{\mathcal{R}^{n+}}\prod_{i=1}^{M_r} SNR^{-\frac{\mathit{b}}{L}(1-\alpha_{1i})^+} \prod_{i=1}^{M_r}SNR^{-(2i-1+M_t-M_r)\alpha_{1i}}d\boldsymbol{\alpha}_1.\nonumber \\
&\doteq&\int_{\mathcal{R}^{n+}}\prod_{i=1}^{M_r}SNR^{-(2i-1+M_t-M_r)\alpha_{1i}-\frac{\textit{b}}{L}(1-\alpha_i)^+}d\boldsymbol{\alpha},  \nonumber \\
&\doteq& SNR^{-\Delta_1} \nonumber
\end{eqnarray}
Again following the arguments of the proof of Theorem 4 in
\cite{Tse}, we have
\begin{eqnarray}
\Delta_1&=&\inf_{\boldsymbol{\alpha}\in\mathcal{R}^{n+}}\sum_{i=1}^{M_r}(2i-1+M_t-M_r)\alpha_{1i}+\frac{\textit{b}}{L}(1-\alpha_{1i})^+.
\end{eqnarray}
The minimizing $\boldsymbol{\alpha}_1$ can be found as
\begin{equation}
\alpha_{1i}= \left\{ \begin{array}{lll}
              0 & \mbox{if} & \frac{\mathit{b}}{L} < 2i-1+M_t-M_r \vspace{.1in}  \\
              1 & \mbox{if} & \frac{\mathit{b}}{L} \geq 2i-1+M_t-M_r. \\
           \end{array} \right.
\end{equation}
Letting $E[D(\mathbf{H})]\dot{\geq} SNR^{-\Delta^{UB}}$, we have $\Delta^{UB} = L \Delta_1$, and
\begin{eqnarray}
\Delta^{UB}=L\sum_{i=1}^{M_r}\min\left\{\frac{b}{L}, 2i-1+M_t-M_r\right\}.
\end{eqnarray}
Similar arguments can be made for the $M_t<M_r$ case, completing the
proof.

\section{Proof of Lemma \ref{l:LP_eqall}}\label{A:LP_eqall}

Let $\mathbf{t^*}$ be the optimal channel allocation vector and
$\mathbf{r}^*$ be the optimal multiplexing gain vector for $n$
layers. For any $\varepsilon>0$ we can find $\mathbf{\tilde{t}}$
with $\tilde{t}_i\in \mathbb{Q}$ and $\sum_{i=1}^n \tilde{t}_i=1$
where $|t^*_i-\tilde{t}_i|<\varepsilon$. Let
$\tilde{t}_i=\gamma_i/\rho_i$ where $\gamma_i\in\mathbb{Z},
\rho_i\in \mathbb{Z}$ and $\theta=LCM(\rho_1,\ldots, \rho_n)$ is the
least common multiple of $\rho_1,\ldots,\rho_n$. Now consider the
channel allocation $\mathbf{\hat{t}}=[1/\theta,\ldots,1/\theta]^T$,
which divides the channel into $\theta$ equal portions and the
multiplexing gain vector
\begin{eqnarray}
\mathbf{\hat{r}}&=&[\underbrace{r^*_1,\ldots,r^*_1}_{\theta
\tilde{t}_1 \mbox{ times}},\underbrace{r^*_2,\ldots,r^*_2}_{\theta
\tilde{t}_2 \mbox{
times}},\ldots,\underbrace{r^*_n,\ldots,r^*_n}_{\theta
\tilde{t}_n\mbox{ times}}]^T
\end{eqnarray}
Due to the continuity of the outage probability and the
distortion-rate function, this allocation which consists of $\theta
n$ layers achieves a distortion exponent arbitrarily close to the
$n$-layer optimal one as $\varepsilon\rightarrow 0$.

Note that $\{\Delta_n^{LS}\}_{n=1}^{\infty}$ is a non-decreasing
sequence since with $n$ layers it is always possible to assign
$t_n=0$ and achieve the optimal performance of $n-1$ layers. On the
other hand, using Theorem \ref{t:upperbound}, it is easy to see that
$\{\Delta_n^{LS}\}$ is upper bounded by $d^*(0)$, hence its limit
exists. We denote this limit by $\Delta^{LS}$. If we define
$\hat{\Delta}^{LS}_n$ as the distortion exponent of $n$-layer LS
with equal channel allocation, we have $\hat{\Delta}^{LS}_n\leq
\Delta^{LS}_n$. On the other hand, using the above arguments, for
any $n$ there exists $m\geq n$ such that $\hat{\Delta}^{LS}_m\geq
\Delta^{LS}_n$. Thus we conclude that
\[ \lim_{n\rightarrow \infty} \hat{\Delta}^{LS}_n = \lim_{n\rightarrow \infty} \Delta^{LS}_n = \Delta^{LS}\]
Consequently, in the limit of infinite layers, it is sufficient to
consider only the channel allocations that divide the channel
equally among the layers.

\section{Proof of Theorem \ref{delta_LS2}}\label{APP:LS2}

We will use geometric arguments to prove the theorem. Using Lemma
\ref{l:LP_eqall}, we assume equal channel allocation, that is,
$\mathbf{t}=[\frac{1}{n}, \ldots, \frac{1}{n}]$. We start with the
following lemma.

\begin{lem}\label{lemma_LS1}
Let $\mathit{l}$ be a line with the equation $y=-\alpha(x-M)$ for
some $\alpha >0$ and $M>0$ and let $\mathit{l_i}$ for $i=1,\ldots,n$
be the set of lines defined recursively from $n$ to $1$ as
$y=(\mathit{b}/n)x+d_{i+1}$, where $\mathit{b}>0$, $d_{n+1}=0$ and
$d_i$ is the $y-$component of the intersection of $\mathit{l_i}$
with $\mathit{l}$. Then we have
\begin{equation}
d_1=M\alpha\left[1-\left(\frac{\alpha}{\alpha+\mathit{b}/n}\right)^{n}\right].
\end{equation}
with
\begin{equation}
\lim_{n\rightarrow\infty}d_1=M\alpha\left(1-e^{-\mathit{b}/\alpha}\right).
\end{equation}\vspace{.01in}
\end{lem}

\begin{proof}
If we solve for the intersection points sequentially we easily find
\begin{equation}
d_{k}-d_{k+1}=M\frac{\mathit{b}}{n}\left(\frac{\alpha}{\alpha+\mathit{b}/n}\right)^{n-k+1},
\end{equation}
for $k=1,\ldots,n$, where $d_{n+1}=0$. Summing up these terms, we
get
\begin{equation}\label{d_k}
d_k=M\alpha\left[1-\left(\frac{\alpha}{\alpha+\mathit{b}/n}\right)^{n-k+1}\right].
\end{equation}
\end{proof}

\begin{figure}
\centering
\includegraphics[width=3.5 in]{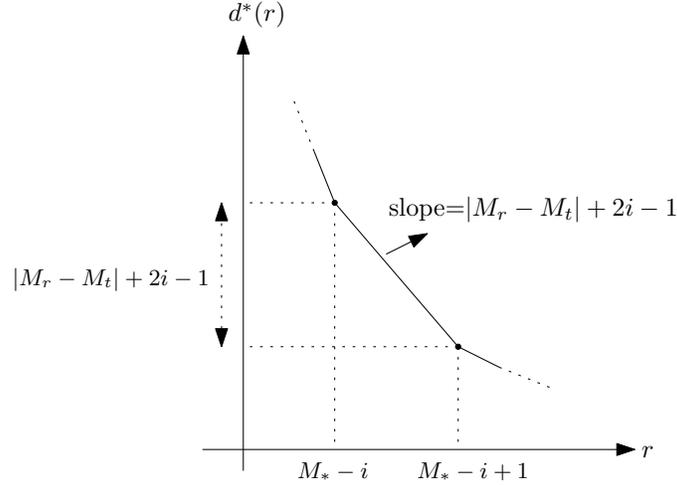}
\caption{DMT curve of an $M_t\times M_r$ MIMO
system is composed of $M_*=\min\{M_t,M_r\}$ line segments, of which
the $i$'th one is shown in the figure.} \label{line_segment}
\end{figure}

In the case of a DMT curve composed of a
single line segment, i.e., $M_*=1$, using Lemma \ref{lemma_LS1} we
can find the distortion exponent in the limit of infinite layers by
letting $M=1$ and $\alpha=M^*$. However, for a general $M_t\times
M_r$ MIMO system the tradeoff curve is composed of $M_*$ line
segments where the $i$th segment has slope $|M_r-M_t|+2i-1$, and
abscissae of the end points $M_*-i$ and $M_*-i+1$ as in Fig.
\ref{line_segment}. In this case, we should consider climbing on
each line segment separately, one after another in the manner
described in Lemma \ref{lemma_LS1} and illustrated in Fig.
\ref{MIMO_stair}. Then, each break point of the
DMT curve corresponds to a threshold on
$\textit{b}$, such that it is possible to climb beyond a break point
only if $\textit{b}$ is larger than the corresponding threshold.

%---------------------------
%---------------------------

Now let $M=M_*-i+1$, $\alpha=|M_r-M_t|+2i-1$ in Lemma
\ref{lemma_LS1} and in the limit of $n\rightarrow\infty$, let $k_in$
be the number of lines with slopes $\mathit{b}/n$ such that we have
$d_n=|M_r-M_t|+2i-1$. Using the limiting form of Eqn. (\ref{d_k}) we
can find that
\begin{equation}
k_i =
\frac{|M_r-M_t|+2i-1}{\mathit{b}}\ln\left(\frac{M_*-i+1}{M_*-i}\right).
\end{equation}
This gives us the proportion of the lines that climb up the $p$th
segment of the DMT curve. In the general
MIMO case, to be able to go up exactly to the $p$th line segment, we
need to have $\sum_{j=1}^{p-1}k_j<1\leq\sum_{j=1}^{p}k_j$. This is
equivalent to the requirement $c_{p-1}< \mathit{b} \leq c_p$ in the
theorem.

To climb up each line segment, we need $k_in$ lines (layers) for
$i=1,\ldots,p-1$, and for the last segment we have
$(1-\sum_{j=1}^{p-1}k_j)n$ lines, which gives us an extra ascent of
\[(M_*-p+1)(|M_r-M_t|+2p-1)(1-e^{-\frac{bk_p}{|M_r-M_t|+2p-1}})\] on
the tradeoff curve. Hence the optimal distortion exponent, i.e., the
total ascent on the DMT curve, depends
on the bandwidth ratio and is given by Theorem \ref{delta_LS2}.

\section{Proof of Lemma \ref{l:HLS2}}\label{APP:HLS2}

As in the proof of Theorem \ref{t:upperbound} in Appendix I, we let
$\lambda_i=SNR^{-\alpha_i}$ and $\bar{\lambda}_i=SNR^{-\beta_i}$ for
$i=1,\ldots, M_*$. The probability densities of
$\boldsymbol{\lambda}$ and $\boldsymbol{\bar{\lambda}}$ and their
exponents $\boldsymbol{\alpha}$ and $\boldsymbol{\beta}$ are given
in Appendix I. Note that since $\mathbf{\bar{H}}$ is a submatrix of
$\mathbf{H}$, $\boldsymbol{\lambda}$ and
$\boldsymbol{\bar{\lambda}}$ as well as $\boldsymbol{\alpha}$ and
$\boldsymbol{\beta}$ are correlated. Let $p(\boldsymbol{\alpha,
\beta})$ be the joint probability density of $\boldsymbol{\alpha}$
and $\boldsymbol{\beta}$. If $M_t \leq M_r$, $\mathbf{H}$ and
$\mathbf{\bar{H}}$ coincide and $\boldsymbol{\lambda} =
\boldsymbol{\bar{\lambda}}$, $\boldsymbol{\alpha}=
\boldsymbol{\beta}$.

We can write
\begin{eqnarray}\label{exp_MMSE}
\int_{\alpha
\in\mathcal{A}^c}\frac{1}{M_*}\sum_{i=1}^{M_*}\frac{1}{1+\frac{SNR}
{M_*}\bar{\lambda}_i}p(\boldsymbol{\lambda})d\boldsymbol{\lambda} &
\doteq & \int_{\alpha \in
\mathcal{A}^c}\sum_{i=1}^{M_*}SNR^{-(1-\beta_i)^+}p(\boldsymbol{\alpha})d\boldsymbol{\alpha},
\end{eqnarray}
\begin{eqnarray}
&\doteq& \int_{\alpha \in\mathcal{A}^c}SNR^{-(1-\beta_{max})^+}
p(\boldsymbol{\alpha})d\boldsymbol{\alpha}, \\
&\doteq& \int_{\alpha \in \mathcal{A}^c}SNR^{-(1-\beta_{max})^+}
\int_{\beta} p(\boldsymbol{\alpha}, \boldsymbol{\beta})
d\boldsymbol{\beta} d\boldsymbol{\alpha},\\
&\doteq& \int_{\beta}SNR^{-(1-\beta_{max})^+} \int_{\alpha \in \mathcal{A}^c} p(\boldsymbol{\alpha}, \boldsymbol{\beta}) d\boldsymbol{\alpha} d\boldsymbol{\beta},\\
&\leq& \int_{\beta}SNR^{-(1-\beta_{max})^+} p(\boldsymbol{\beta}) d\boldsymbol{\beta} ,\\
&\doteq& SNR^{-\mu},
\end{eqnarray}
where
\begin{eqnarray}
\mu &=& \inf _{\beta\in \mathbb{R}^{M_*+}} (1-\beta_{max})^+ +
\sum_{i=1}^{M_*}(2i-1)\beta_i.
\end{eqnarray}

The minimizing $\tilde{\boldsymbol{\beta}}$ satisfies
$\tilde{\beta}_1\in [0,1]$ and
$\tilde{\beta}_2=\cdots=\tilde{\beta}_{M_*}=0$, and we have $\mu=1$.

\section{Proof of Theorem \ref{t:MIMO_DMT}}\label{APP:MIMO_DMT}

The mutual information between
$\mathbf{Y}_k$ and $\mathbf{X}_k$ defined in (\ref{eq:channel}) can be written as
\begin{eqnarray}
\mathcal{I}(\mathbf{Y}_k;\mathbf{X}_k) &=&
\mathcal{I}(\mathbf{Y}_k;\mathbf{X}_k,\mathbf{\bar{X}}_{k+1}) -
\mathcal{I}(\mathbf{Y}_k;\mathbf{\bar{X}}_{k+1}|\mathbf{X}_k), \\
&=&\log \det \left(\mathbf{I} +
\frac{\overline{SNR}_k}{M_t}\mathbf{HH}^\dag\right) - \log \det
\left(\mathbf{I} +
\frac{\overline{SNR}_{k+1}}{M_t}\mathbf{HH}^\dag \right),\nonumber \\
&=& \log \frac{\det \left(\mathbf{I}+
\frac{\overline{SNR}_k}{M_t}\mathbf{HH}^\dag\right) }{\det
\left(\mathbf{I} + \frac{\overline{SNR}_{k+1}}{M_t}\mathbf{HH}^\dag
\right)}. \label{mutual_BS}
\end{eqnarray}

For layers $k=1,\ldots,n-1$, and the multiplexing gain vector
$\mathbf{r}$ we have
\begin{eqnarray}
P_{out}^k &=& Pr \left\{\mathbf{H}: \log \frac{\det
\left(\mathbf{I}+ \frac{\overline{SNR}_k}{M_t} \mathbf{HH}^\dag
\right)} {\det \left(\mathbf{I} + \frac{\overline{SNR}_{k+1}}{M_t}
\mathbf{HH}^\dag \right)}
< r_k\log SNR\right\} \nonumber  \label{cond_half} \\
&=& Pr \left\{\mathbf{H}: \frac{\prod_{i=1}^{M_*}(1+
\frac{\overline{SNR}_k}{M_t} \lambda_i)} {\prod_{i=1}^{M_*} (1+
\frac{\overline{SNR}_{k+1}}{M_t} \lambda_i)} < SNR^{r_k}\right\},
\label{cond1}
\end{eqnarray}
and
\begin{equation}\label{cond2}
P_{out}^n = \left\{ \mathbf{H}: \prod_{i=1}^{M_*}\left(1+
\frac{\overline{SNR}_n}{M_t} \lambda_i \right) < SNR^{r_n} \right\}.
\end{equation}
where $\lambda_1\leq \lambda_2\leq \cdots \leq \lambda_{M_*}$ are
the eigenvalues of $\mathbf{HH}^\dag$ ($\mathbf{H^\dag H}$) for
$M_t\geq M_r$ ($M_t< M_r$). Let $\lambda_i=SNR^{-\alpha_i}$. Then
for the power allocation in (\ref{pow_alloc_MIMO}), conditions in
Eqn. (\ref{cond1}) and Eqn. (\ref{cond2}) are, respectively,
equivalent to
\begin{eqnarray}
\sum_{i=1}^{M_*}(1-r_1-\cdots-r_{k-1}-\epsilon_{k-1}-\alpha_i)^+ -
\sum_{i=1}^{M_*}(1-r_1-\cdots-r_k-\epsilon_k-\alpha_i)^+ < r_k,
\nonumber
\end{eqnarray}
and
\begin{equation}
\sum_{i=1}^{M_*}(1-r_1-\cdots-r_{n-1}-\epsilon_{n-1}-\alpha_i)^+ <
r_n. \nonumber
\end{equation}

Using Laplace's method and following the similar arguments as in the
proof of Theorem 4 in \cite{Tse} we show that, for $k=1,\ldots,n$,
\begin{eqnarray}
P_{out}^k&\doteq&SNR^{-d_k},
\end{eqnarray}
where
\begin{eqnarray}
d_{k} &=& \inf_{\boldsymbol{\alpha}\in
\tilde{A}_k}\sum_{i=1}^{M_*}\left(|M_t-M_r|+2i-1\right)\alpha_i.
\end{eqnarray}
For $k=1,\ldots,n-1$

\begin{eqnarray}
\begin{array}{ll}
\tilde{A}_k=\bigg\{&\boldsymbol{\alpha}= [\alpha_1,\ldots,\alpha_{M_*}]\in
\mathbb{R}^{M_*+}:\alpha_1\geq\cdots\geq\alpha_{M_*}\geq 0,\\
&\sum_{i=1}^{M_*}(1-r_1-\cdots-r_{k-1}-\epsilon_{k-1}-\alpha_i)^+
-\sum_{i=1}^{M_*}(1-r_1-\cdots-r_k-\epsilon_k-\alpha_i)^+ < r_k
\bigg\}.
\end{array}\nonumber
\end{eqnarray}
while
\begin{equation}
\begin{array}{ll}
\tilde{A}_n = \{&\boldsymbol{\alpha} = [\alpha_1,\ldots,\alpha_{M_*}]\in
\mathbb{R}^{M_*+}:\alpha_1\geq\cdots\geq\alpha_{M_*}\geq 0,\\
&\sum_{i=1}^{M_*}\left(1-r_1-\cdots-r_{n-1}-\epsilon_{n-1}-\alpha_i\right)^+
< r_n \}.
\end{array}\nonumber
\end{equation}\vspace{.01in}

The minimizing $\boldsymbol{\tilde{\alpha}}$ for each layer can be
explicitly found as
\begin{eqnarray}
\tilde{\alpha}_i &=& 1-r_1-\cdots-r_{k-1}-\epsilon_{k-1}, \mbox{ for
 }i=1,\ldots,M_*-1, \nonumber
\end{eqnarray}
and
\begin{eqnarray}
\tilde{\alpha}_{M_*} &=& 1-r_1-\cdots-r_{k-1}-r_k-\epsilon_{k-1}.
\nonumber
\end{eqnarray}
Letting $\epsilon_k\rightarrow 0$ for $k=1,\ldots,n-1$, we have
\begin{eqnarray}
d_k &= M^*M_*(1-r_1-\cdots-r_{k-1}) - (M^*+M_*-1)r_k .
\end{eqnarray}

Note that the constraint $\sum_{i=1}^n r_i \leq 1$ makes the
sequence $\{1-r_1-\cdots-r_k\}_{k=1}^n$ decreasing and greater than
zero. Thus $P_{out}^k$ constitutes an increasing sequence.
Therefore, using (\ref{d:B_k}) and (\ref{pout_bar}), in the high SNR
regime we have $\bar{P}_{out}^k \doteq P_{out}^k$, and $d_{sd}(r_k)
= d_k$.

\section{Proof of Theorem \ref{t:MIMO_DE}}\label{APP:MIMO_DE}

Using the formulation of $\Delta_n^{BS}$ in (\ref{BS_ED_hSNR}) and
successive decoding diversity gains of the proposed power allocation
in Theorem \ref{t:MIMO_DMT}, we find the multiplexing gain
allocation that results in equal SNR exponents for all the terms in
(\ref{BS_ED_hSNR}).

We first consider the case $\textit{b}\geq (M_t-1)(M_r-1)$. Let
\begin{equation}
\eta_0 = \frac{\textit{b}- (M_t-1)(M_r-1)}{M_t+M_r-1} \geq0.
\end{equation}
For $0\leq \eta_0 < 1$, we set
\begin{eqnarray}\label{BS_mux1}
r_1 &=&\frac{M_tM_r(1-\eta_0)}{M_tM_r-\textit{b}\eta_0^n},\\ % \nonumber \\
r_i &=& \eta_0^{i-1}r_1, \mbox{ for } i=2,\ldots,n. \label{BS_mux2}%\nonumber
\end{eqnarray}
If $\eta_0 \geq 1$, we set
\begin{eqnarray}\label{BS_mux2}
r_1 = \cdots = r_n = \frac{M_tM_r}{nM_tM_r+M_t+M_r-1}. % \nonumber \\
\end{eqnarray}

Next, we show that the above multiplexing gain assignment satisfies
the constraint $\sum_{i=1}^{n}r_i\leq1$. For $\eta_0 <1$, $b \leq
M_tM_r$ and
\begin{eqnarray}
r_1+\cdots+r_n &=& r_1\left(1+\eta_0+\cdots+\eta_0^{n-1} \right), \nonumber \\
 &=& \frac{M_tM_r(1-\eta_0^n)}{M_tM_r-\textit{b}\eta_0^n}, \nonumber \\
 &\leq & 1.
\end{eqnarray}
On the other hand, when $\eta_0 \geq1$, we have $\sum_{i=1}^n
r_i=\frac{nM_tM_r}{nM_tM_r+M_t+M_r-1}<1$.

Then the corresponding distortion exponent can be found as
\begin{equation}
\Delta^{BS}_n  = \left\{ \begin{array}{lll}
              b\frac{M_tM_r(1-\eta_0^n)}{M_tM_r-\textit{b}\eta_0^n} & \mbox{if} & (M_t-1)(M_r-1)\leq b<M_tM_r, \vspace{.02in} \\
              \frac{n(M_tM_r)^2}{nM_tM_r+M_t+M_r-1} & \mbox{if} & b \geq M_tM_r. \\
           \end{array} \right.
\end{equation}

For $(M_t-k-1)(M_r-k-1) \leq b< (M_t-k)(M_r-k)$, $k=1,\ldots,M_*-1$,
we can consider the $(M_t-k) \times (M_r-k)$ antenna system and
following the same steps as above, we obtain a distortion exponent
of
\begin{equation}
b\frac{(M_t-k)(M_r-k)(1-\eta_k^n)}{(M_t-k)(M_r-k)-b\eta_k^n},
\nonumber
\end{equation}
where $\eta_k$ is defined in (\ref{eta_k}).

In the limit of infinite layers, it is possible to prove that this
distortion exponent converges to the following.

\begin{eqnarray}
\Delta^{BS} = \lim_{n\rightarrow\infty} \Delta^{BS}_n = \left\{
\begin{array}{lll}
                \textit{b} & \mbox{if} & 0 \leq b < M_tM_r, \vspace{.02in} \\
                M_tM_r & \mbox{if} & b\geq M_tM_r. \\
           \end{array} \right.
\end{eqnarray}

\section{Proof of Theorem \ref{t:block_DE}}\label{APP:block_DE}

We transmit codewords of each layer across all fading blocks,
which means that $P_{out}^k$ in Eqn. (\ref{cond_half}) becomes
\begin{eqnarray}
P_{out}^k &=& Pr \left\{(\mathbf{H}_1,\ldots,\mathbf{H}_L):
\frac{1}{L}\sum_{i=1}^{L}\log \frac{\det \left(\mathbf{I}+
\frac{\overline{SNR}_k}{M_t} \mathbf{H}_i\mathbf{H}_i^\dag \right)}
{\det \left(\mathbf{I} + \frac{\overline{SNR}_{k+1}}{M_t}
\mathbf{H}_i\mathbf{H}_i^\dag \right)} < r_k\log SNR\right\},
\nonumber
\end{eqnarray}
for $k=1,\ldots,n-1$, and $P_{out}^n$ can be defined similarly.

For the power allocation in (\ref{pow_alloc_BF}), above outage event
is equivalent to
\begin{eqnarray}
\sum_{j=1}^{L} \sum_{i=1}^{M_*}(1-Lr_1-\cdots-Lr_{k-1}-L\epsilon_{k-1}-\alpha_{j,i})^+ -
(1-Lr_1-\cdots- Lr_k-L\epsilon_k-\alpha_{j,i})^+ < Lr_k.
\nonumber
\end{eqnarray}

Following the same steps as in the proof of Theorem \ref{t:MIMO_DMT}, we have
\begin{eqnarray}
P_{out}^k& \doteq &SNR^{-d_k},
\end{eqnarray}
where
\begin{eqnarray}
d_{k} &=& \inf_{\boldsymbol{\alpha}\in
\tilde{A}_k} \sum_{j=1}^{L} \sum_{i=1}^{M_*} \left(|M_t-M_r|+2i-1\right)\alpha_{j,i}.
\end{eqnarray}
For $k=1,\ldots,n-1$
\begin{eqnarray}
\begin{array}{ll}
\tilde{A}_k=\bigg\{&\boldsymbol{\alpha}= [\alpha_1,\ldots,\alpha_{LM_*}]\in
\mathbb{R}^{LM_*+}:\alpha_{j,1}\geq\cdots\geq\alpha_{j,M_*}\geq 0 \mbox{ for } j=1,\ldots,L\\
& \sum_{j=1}^{L} \sum_{i=1}^{M_*}(1-Lr_1-\cdots-Lr_{k-1}-L\epsilon_{k-1}-\alpha_{j,i})^+  \\
& -(1-Lr_1-\cdots-Lr_k-L\epsilon_k-\alpha_{j,i})^+ < Lr_k \bigg\}.
\end{array}\nonumber
\end{eqnarray}
%while
%\begin{equation}
%\begin{array}{ll}
%\tilde{A}_n = \{&\boldsymbol{\alpha} = [\alpha_1,\ldots,\alpha_{M_*}]\in
%\mathbb{R}^{M_*+}:\alpha_1\geq\cdots\geq\alpha_{M_*}\geq 0,\\
%&\sum_{i=1}^{M_*}\left(1-r_1-\cdots-r_{n-1}-\epsilon_{n-1}-\alpha_i\right)^+
%< r_n \}.
%\end{array}\nonumber
%\end{equation}\vspace{.01in}

The minimizing $\boldsymbol{\tilde{\alpha}}$ for each layer can be
explicitly found as
\begin{eqnarray}
\tilde{\alpha}_{j,i} &=& 1-Lr_1-\cdots-Lr_{k-1}-L\epsilon_{k-1}, \mbox{ for
 } j=1,\ldots,L, \mbox{ and } i=1,\ldots,M_*-1, \nonumber
\end{eqnarray}
and
\begin{eqnarray}
\tilde{\alpha}_{j,M_*} &=& 1-Lr_1-\cdots-Lr_{k-1}-r_k-L\epsilon_{k-1}, \mbox{ for
 } j=1,\ldots,L.
\nonumber
\end{eqnarray}
Then, letting $\epsilon_{n-1} \rightarrow 0$, the diversity gain $d_k$ is obtained as
\begin{eqnarray}
d_k &= L \bigg[ M^*M_*(1-Lr_1-\cdots-Lr_{k-1}) - (M^*+M_*-1)r_k \bigg].
\end{eqnarray}

We skip the rest of the proof as it closely resembles the proof of Theorem \ref{t:MIMO_DE} in Appendix \ref{APP:MIMO_DE}.

\end{document}